\documentclass[11pt,a4paper]{article}
\usepackage{jheppub}

\usepackage{hyperref}
\usepackage{comment}
\usepackage{color}

\usepackage{setspace}
\usepackage{amsmath,amssymb,amsfonts,amsthm,mathrsfs}
\usepackage{cancel}
\usepackage{graphicx}
\usepackage{enumerate} 

\def\be{\begin{equation}}
\def\ee{\end{equation}}
\def\ba{\begin{eqnarray}}
\def\ea{\end{eqnarray}}
\def\bi{\begin{itemize}}
\def\ei{\end{itemize}}

\def\ket{\rangle}

\def\zb{\bar{z}}
\def\wb{\bar{w}}
\def\xh{\hat{x}}

\def\L{\mathcal{L}}

\def\w{\omega}

\def\hard{\text{hard}}
\def\soft{\text{soft}}

\def\R{\mathcal{R}}

\def\D{\mathcal{D}}
\def\Pit{\tilde{\Pi}}

\def\extra{\text{extra}}
\def\H{\mathcal{H}}
\def\Gkin{\Gamma^{\textrm{kin}}}
\def\Okin{\Omega^{\textrm{kin}}}
\def\sigmao{\mathring{\sigma}}

\def\Ch{\widehat{C}}
\def\Nzeroh{\widehat{N}^0}
\def\Gh{\widehat{G}}
\def\Noneh{\widehat{N}^1}
\def\Dh{\widehat{D}}
\def\Th{\widehat{T}}
\def\Nh{\widehat{N}}
\def\Ph{\hat{P}}
\def\Jh{\hat{J}}
\def\Qh{\hat{Q}}
\def\Dcalh{\widehat{\D}}

\def\F{\mathcal{F}}
\def\K{\mathcal{K}}

\newcommand\Nzero{\overset{\scriptscriptstyle 0}{N}\vphantom{N}}
\newcommand\None{\overset{\scriptscriptstyle 1}{N}\vphantom{N}}
\newcommand\Gpsi{\overset{\scriptscriptstyle \psi}{\Gamma}\vphantom{\Gamma}}

\newcommand\Nbzero{\overset{\scriptscriptstyle 0}{\mathcal{N} }\vphantom{\mathcal{N} }}
\newcommand\Nbone{\overset{\scriptscriptstyle 1}{\mathcal{N} }\vphantom{\mathcal{N} }}

\title{BMS Algebra, Double Soft Theorems, and All That}

\author[a]{Miguel Campiglia} 
\author[b]{Alok Laddha} 
\affiliation[a]{Instituto de F\'isica, Facultad  de  Ciencias, 
Ig\'ua  4225,  Montevideo,  Uruguay}
\affiliation[b]{Chennai Mathematical Institute,  SIPCOT IT Park,
Siruseri 603103, India}
\emailAdd{campi@fisica.edu.uy}
\emailAdd{aladdha@cmi.ac.in}

\abstract{
The Lie algebra generated by supertranslation and superrotation vector fields at null infinity, known as the extended  Bondi–van der Burg–Metzner–Sachs (eBMS) algebra is expected to be a symmetry algebra of the quantum gravity S-matrix.  However, the algebra of commutators of the quantized eBMS charges has been a thorny issue in the literature. On the one hand, recent developments in celestial holography point towards a symmetry algebra which is a closed Lie algebra with no central extension or anomaly; and on the other hand, work of Distler, Flauger and Horn has shown that when these charges are quantized at null infinity, the commutator of a supertranslation and a superrotation charge does not close into a supertranslation but gets deformed by a 2-cocycle term, which is consistent with the original proposal of Barnich and Troessaert. 

In this paper, we revisit this issue in light of recent developments in the classical understanding of superrotation charges. We show that,  for extended BMS symmetries, a phase space at null infinity is an extension of hitherto considered phase spaces which also includes a mode associated to the spin-memory and its conjugate partner. We also show that  for holomorphic vector fields on the celestial plane,  quantization of the eBMS charges in the new phase space leads to an algebra which closes without a 2-cocycle. The degenerate vacua are labelled by the soft news and a Schawarzian mode which corresponds to deformations of the celestial metric by superrotations. The closed eBMS  quantum algebra may also lead to a convergence between two manifestations of asymptotic symmetries, one via asymptotic quantization at null infinity and the other through celestial holography.}

\begin{document}

\maketitle

\tableofcontents

\section{Introduction}

The Bondi–van der Burg–Metzner–Sachs (BMS) symmetry algebra of General Relativity \cite{bms1,bms2} has received renewed attention over the past decade thanks to seminal results by Strominger et. al. \cite{strom1,stromST,stromSR,stromzhibo}.  The symmetries of classical as well as perturbative quantum gravitational scattering (which may include massive scattering particles) includes a universal BMS symmetry group which acts on all the boundaries of spacetime including time-like infinity. At  quantum level, the existence of BMS symmetries imply a highly intricate structure of the algebra of boundary observables which can be used to formulate flat space holography in terms of such algebra of observables \cite{suvrat-flat}, paralleling  the conception of (far more developed) holography in asymptotically AdS spacetimes \cite{suvrat-ads}.

In fact, it is by now becoming increasingly clear  that the BMS symmetry algebra, which is a semi-direct product of supertranslations and celestial Lorentz transformations, gets enlarged into  two possible infinite dimensional extensions in which the Lorentz algebra is embedded, either  the algebra of meromorphic vector fields  \cite{btebms} or  the algebra of smooth vector fields \cite{clgbms} on the celestial sphere.  These two extensions are known in the literature as extended BMS  (eBMS) and generalised BMS (gBMS)  respectively. We will refer to both classes of celestial vector fields, that are either meromorphic or smooth, as superrotation vector fields. 

At the level of the algebra generated by supertranslation and superrotation vector fields, both extensions of the BMS algebra form an infinite dimensional Lie algebra. Representation of eBMS algebra is an active area of investigation. Coadjoint orbits of (e/g)BMS algebra have been studied in \cite{freidelorbit}, \cite{barnichruzz}, paving the way for its quantization via the orbit method \cite{kirilov}. In  the seminal work \cite{cordova}, it was shown that gBMS algebra is faithfully represented in a unitary conformal field theory in Minkowski spacetime as the algebra of so called light-ray operators on lightlike boundaries of finite regions.

However the situation is more intricate in gravity at null infinity. As remarkably shown by Barnich and Troessaert (BT)  \cite{btchargealgebra}, the eBMS charge algebra at null infinity does not close and the non-closure term is in fact not a central extension but linear in the asymptotic gravitational field. Technically, the extension is a 2-cocycle \cite{jackiw}. That is, it is a closed 2-form on the space of eBMS generators.\footnote{A standard example of 2-cocycle occurs in anomalous gauge theories. That is, the commutator of two gauge generators in an anomalous gauge theory is a 2-cocycle.} This result was confirmed in a beautiful paper  by  Distler, Flauger, and Horn (DFH) \cite{dfh} by evaluating the commutator of the quantum eBMS charges on the Fock space of asymptotic gravitons. Their result complemented an earlier analysis in \cite{aar} which showed how Ward identities associated to e/gBMS generators in the so-called shifted vacuum were consistent with double soft graviton theorem. 
It was also proved in \cite{dfh} that the eBMS algebra with the BT extension is consistent with the so called consecutive double soft graviton theorem for tree-level amplitudes.  More in detail, the authors showed that the right hand side of the commutator of two consecutive double soft theorems up to sub-sub leading order  equals the matrix element of the 2-cocycle extended eBMS algebra.

However this results appears to be in tension with the following: 
\begin{itemize}

\item As a consequence of Jacobi identity, we expect the commutators of two symmetries of the S-matrix to be a symmetry of the S-matrix. 
A centrally  extended algebra is consistent with this basic fact, but the presence of a 2-cocycle obstructs the interpretation of eBMS algebra as a symmetry algebra of  quantum gravity. In other words, eBMS symmetry appears to be anomalous even for tree-level S matrix!

\item Recent developments in celestial holography  have led to a novel realization of the eBMS algebra  \cite{ebmsope, shamik, alfredo} from  conformally soft graviton theorems \cite{confsoft1,confsoft2,confsoft3}. This realisation provides a strong evidence that in the infrared sector, flat-space holography is captured by  a conformal field theory (CFT) with a supertranslation current that generates an abelian ideal and a holographic stress tensor that generates a Witt algebra with vanishing central charge.\\
In \cite{ebmsope, alfredo}, the celestial  eBMS algebra was derived from the operator product expansion (OPE) involving supertranslation current and celestial stress tensor, which are 
 the supertranslation and superrotation soft charges in the celestial basis. 
This OPE is in turn obtained from conformal soft graviton theorems. In particular, the OPE between the supertranslation current $P(z, \zb)$ and the (shadow of)  superrotation soft mode ${\cal T}_{ww}$ has a double pole $\frac{1}{(z-w)^{2}}$ \cite{ebmsope,shamik}. From this perspective, the appearance of the BT cocycle anomaly is rather mysterious since it corresponds to a forth order pole in the corresponding OPE \cite{dfh}. In fact, the presence of such a term hints at a possibility that the supertranslation current is not a primary (we refer to section 5 of \cite{dfh} for a detailed discussion on this issue).  
\end{itemize}

In this paper we take the first steps in resolving these issues. More in detail, we show that for a class of superrotation vector fields, the ``improved" superrotation charges derived in \cite{compere,cp} lead to a closed eBMS quantum algebra with no cocycle anomaly.  In particular, we consider superrotations generated by holomorphic (or anti-holomorphic) vector fields on the celestial sphere which only have singularity at the North pole. This is equivalent to de-compactifying the celestial sphere to a plane and consider superrotations generated by holomorphic (or anti-holomorphic) as opposed to meromorphic vector fields.
 
Our starting point is the classical analysis of gBMS charges initiated in \cite{compere,cp, wbms}. In \cite{compere} it was shown how the asymptotic phase space of gravity can be consistently renormalized in order to obtain canonical gBMS charges at null infinity. This led to several corrections in the original charge expressions  \cite{stromSR,clpilot}, in particular due to their non-trivial dependence on the Geroch tensor \cite{geroch}.  The charge algebra of \cite{compere}, however, still displayed a 2-cocycle. In \cite{cp} it was noticed that the 2-cocycle could be eliminated by a total derivative term in the superrotation charge, and the corresponding  correction on the gravitational symplectic structure was derived. These anomaly-free  gBMS charges were then re-derived from other approaches \cite{lambdabms,comperenichols}, thus giving confidence on the proposal. Whereas these results were obtained for generalized BMS charges, they apply equally well to their extended version, leading to an anomaly-free eBMS charge algebra.

It is not clear, however, if the above classical results carry over to quantum theory, and if so how they relate to the DFH quantum analysis. A major obstacle arises because the extended radiative phase of \cite{cp} includes soft modes associated to superrotation symmetry which are not all independent but are subject to certain non-linear constraints. This complicates the task of isolating the physical degrees of freedom and computing the corresponding Poisson brackets. Without a complete understanding of the physical phase space,  the quantization and the connection with the results of \cite{dfh} has remained out of reach. 

The analysis of \cite{cp} also relied on an extension of the  radiative phase space of GR which did not  explicitly include the supertranslation modes.  Hence this extension remained disparate from the phase space derived by He, Lysov, Mitra and Strominger (HLMS) \cite{stromST} in which a \emph{physical}  phase space that included supertranslation soft modes was derived. 

We start by filling both  gaps in the present paper.  That is, we derive a  phase space at null infinity which we refer to as $\Gamma$ and show that it contains supertranslation modes, superrotation modes and their conjugate partners. However the phase space we obtain does not contain all superrotation modes.  A generic superrotation (either meromorphic or smooth) deforms the metric on the celestial sphere. Quantizing the celestial metric at ${\cal I}$ is subtle and we do not attempt to do so in this paper. Instead, we consider the (infinite dimensional) algebra of holomorphic (or anti-holomorphic) superrotations which have poles only at infinity in the celestial plane. 

We then show that we can quantize $\Gamma$ and obtain the quantum version of the  supertranslation and the superrotation charges defined in \cite{compere, cp}. The quantum algebra generated by these charges closes without a 2-cocycle  and show how eBMS is a symmetry of tree-level S matrix for holomorphic superrotations. As there is no free lunch, there is a price to pay. Namely, the superrotation soft charge contains in addition to the linear soft operator (which is proportional to the spin memory), a quadratic operator that depends on leading soft news and its conjugate constant shear. This operator has a trivial action on Fock states defined over the trivial vacuum, but has a non-trivial action on graviton states defined with respect to  supertranslation displaced vacua.  

The paper is organised as follows. In section \ref{secone}, we briefly review the radiative phase space derived in \cite{stromST} which contained an explicit parametrization of supertranslation modes. In section \ref{sec3} we review and extend the analysis of \cite{cp} to obtain a phase space $\Gamma$ that explicitly includes supertanslation as well as superrotation modes. In section \ref{sec4} we construct an auxiliary ``kinematical" phase space $\Gkin$ in which these ``soft'' modes are decoupled from the ``hard'' fields such that $\Gamma$ can  be understood as a constrained phase space inside $\Gkin$. This description is used in section \ref{secfour} to obtain Poisson brackets in $\Gamma$ via a Dirac constraint analysis. 
In section \ref{secfive} we quantize the physical phase space and show that the corresponding quantized charges generate a closed eBMS algebra. 
We also show how our results are consistent with earlier results in the literature including those  of \cite{dfh}.  
We conclude in section \ref{sec7} with a summary of results and open questions and offer some preliminary comments on the relationship between quantized eBMS algebra derived in this paper and the eBMS algebra in celestial holography.

As we shall see, our focus on holomorphic rather than meromorphic superrotations  greatly simplifies the analysis. The extension of our work to complete eBMS as well as gBMS  requires further investigation. In both of these cases, the complete radiative phase space should include modes associated to deformation of the metric on the celestial sphere.
In the conclusion section, we briefly comment on some ideas for extending our results to the case where superrotations are generated by meromorphic (or smooth) rather than holomorphic, vector fields on the celestial sphere. \\

\noindent \emph{Conventions:}\\
We work in units such that $32 \pi G =1$. Given a function $f$ on a phase space with symplectic structure $\Omega$, we define its Hamiltonian vector field (HVF) $X_f$  by the condition $\delta f = \Omega(\delta, X_f)$. The PB between two functions $f$ and $g$   is then defined by $\{f, g\} := \Omega(X_g, X_f) = X_g(f) = - X_f(g)$.

\section{Review of the radiative phase space}\label{secone}
In this section, we briefly review the classical definition of gravitational radiative phase space in General Relativity \cite{as81}. We then review one of the proposed extensions  \cite{stromST} in which the soft news and its conjugate partner are included as part of the phase space degrees of freedom. Our review is rather brief and the interested reader is encouraged to consult the cited papers for more details. 

The radiative phase space of General Relativity  at future null infinity ${\cal I}^{+}$ was derived by Ashtekar and Streubel (AS) in \cite{as81}. Although the radiative phase space is gauge invariant and can be defined in a coordinate-independent manner, it can be most easily motivated by analysing the space of solutions to Einstein equations in Bondi gauge close to future null infinity ${\cal I}^{+}$,
\begin{flalign}
ds^{2}\, =\, -\, du^{2}\, -\, 2 du\, dr\, +\, r^{2}\, ( q_{ab}\, +\, \frac{1}{r}\, \sigma_{ab}\, ) dx^{a}dx^{b}\, +\, O(\frac{1}{r^{2}})
\end{flalign}
where $(a,b)$ label coordinates on the celestial sphere (we shall later choose them to be stereographic coordinates), $q_{ab}$ is the unit 2-sphere metric and $\sigma_{ab}=\sigma_{ab}(u,\xh)$ is the shear field associated to the true degrees of freedom of the radiative gravitational field.  We denote by $(u,\xh)$ points on future null infinity, with $u$ the retarded time and $\xh$ the direction on the celestial sphere. 

Ashtekar and Streubel showed that the phase space of General Relativity without matter sources is parametrised by the shear field $\sigma_{ab}(u, \hat{x})$ with the symplectic structure
\be
\Omega_{0}(\delta,\, \delta^{\prime})\, =\, \int_{{\cal I}^{+}}\, du\, d^{2}\hat{x}\, [\, \delta\, \sigma_{ab}\, \delta^{\prime}\, N^{ab}\, -\, \delta\, \leftrightarrow\, \delta^{\prime}\, ]
\ee
where  $N_{ab}\, =\, \partial_{u}\, \sigma_{ab}$ is the news tensor at ${\cal I}^{+}$. The finiteness of $\Omega_0$ is guaranteed by the fall-off conditions on the shear as we approach the two boundaries of ${\cal I}^{+}$.  AS require minimal fall-offs for this to happen, namely 
\be
N_{ab}(u, \hat{x})\, \overset{ u \rightarrow\, \pm\infty}{=}\, O(1/|u|^{1 + \epsilon}) \quad \text{(AS fall-offs)} \label{ASfalloffs}
\ee
 with $\epsilon>0$. However, in order to be able to define superrotation charges in  later sections, we shall require the stronger fall-offs
\be
N_{ab}(u, \hat{x})\, \overset{ u \rightarrow\, \pm\infty}{=}\, O(1/|u|^{2 + \epsilon}) \quad \text{($\Gamma_{0}$ fall-offs)} \label{ourfalloffs}
\ee
with $\epsilon>0$. We will refer to this phase space as $\Gamma_{0}$.\footnote{The discussion in the remainder of this section is however valid for the weaker fall-offs \eqref{ASfalloffs}. It is interesting to note that generic radiative fall-offs are of the form $N_{ab}=O(1/|u|^2)$ (see e.g. \cite{ashoke,biswajit}). This is allowed by \eqref{ASfalloffs} but forbidden in \eqref{ourfalloffs}. Fall-offs of the type \eqref{ourfalloffs} are found when computing gravitational radiation at tree-level and correspond to a  $O(\w^0)$ subleading soft emission. The $N_{ab}=O(1/|u|^2)$ behavior is associated to the so-called tail to the memory effect and corresponds to a $O(\ln \w)$ subleading soft emission \cite{ashoke,biswajit}. We leave the generalisation of our work and a completely satisfactory construction of the radiative phase space that includes  tail to the memory modes for future work. For some progress in this direction we refer the reader to \cite{cl1903,sayali}.}

BMS symmetries, that is supertranslations as well as Lorentz transformations are represented on $\Gamma_{0}$ by their actions on the shear and news tensors. However, only the latter admits a realization in term Poisson brackets, see e.g. \cite{stromST,clgbms}. In order to fully capture the action of supertranslations via Poisson brackets, one needs an enhancement of the radiative phase space. Precisely such an extension  was proposed by He, Lysov, Mitra and Strominger (HLMS) in \cite{stromST}, with a phase space  of the form
\be
\Gamma_{\textrm{HLMS}}\,\subset\, \Gamma_{0}\, \times \Gamma_{\textrm{s}}, \label{GammaHLMS}
\ee
where $\Gamma_{\textrm{s}}$ is the ``soft sector" parametrized by the soft (zero frequency component of) the news and its conjugate which is simply the $u$ independent mode of $\sigma_{ab}$. More in detail, the soft sector is a $2\, \times\, \infty$ dimensional phase space parametrized by,
\be
\begin{array}{lll}
\Nzero(\hat{x})\, : \Nzero_{ab}(\hat{x})\, =\, D_{a}D_{b}\, \Nzero(\hat{x})\\
C(\hat{x})\, :\, C_{ab}\, =\, D_{a}D_{b}\, C(\hat{x}) .
\end{array}
\ee
However as we have not changed the theory, the additional degrees of freedom are not independent but are related to the fields in $\Gamma_{0}$ via,  
\ba
\lim_{u \to -\infty} C_{ab}(u, \hat{x}) & =& D_{a}D_{b}C(\hat{x}),\\
\int du\, N_{ab}(u, \hat{x}) &=& D_{a}D_{b}\, \Nzero(\hat{x}).
\ea
Following \cite{strom1}, we refer to these conditions as the Christodolou-Klainermann (CK) constraints. These are the constraints that define the phase space $\Gamma_{\textrm{HLMS}}$ in \eqref{GammaHLMS}.  Physically they show that the leading soft news (the displacement memory) is associated to a single function on the celestial sphere.\footnote{They can also be understood in terms of vanishing of the so-called dual supertranslation charges, see e.g. \cite{subsub2,pope}.} As \cite{stromST} showed, the CK constraints expressed in this form  are second class constraints and one has to solve them by employing the standard technique of Dirac brackets. 
The resulting brackets are such that the soft degrees of freedom Poisson commute with the finite energy news $N_{ab}(u,\hat{x})$ but they have non-trivial Poisson brackets with the shear field $\sigma_{ab}(u, \hat{x})$. 


\section{Radiative phase space with superrotation modes}\label{sec3}
In this section we review the phase space defined in \cite{cp}, which is an extension of $\Gamma_{0}$ obtained by adding modes associated to superrotations.  The  extension in \cite{cp} was defined for  $\textrm{Diff}(S^{2})$ superrotations and so the  phase space  included smooth deformations of the sphere metric. In this paper however, we restrict  attention to holomorphic vector fields. They leave the 2d metric invariant but act non-trivially on the space of  Geroch tensors \cite{comperelong}. For simplicity, we will work in a conformal frame where the celestial 2d metric is the flat one.

We thus start by considering Bondi coordinates $(r,u,z,\zb)$  such that the $r \to \infty$ expansion of  the spacetime metric takes the form
\be
ds^2  = -2 du dr + 2 r^2  d z d \zb + r ( (\sigma_{zz}(u,z,\zb) + u T_{zz}(z,\zb)) dz dz + c.c.) + \cdots . \label{gzz}
\ee
The metric written above is based on the following conventions and assumptions:
\begin{itemize}
\item We choose the metric on null infinity to be the 2d flat metric, $q_{z \zb} =1, q_{zz}=q_{\zb \zb}=0$. Whereas our analysis readily generalizes to non-flat metrics, the underlying plane topology assumption will play a key simplifying role.\footnote{Strictly speaking, our analysis is only be valid for the case where the celestial surface is a plane rather than a sphere. We will however continue to use the term "celestial sphere" to refer to this surface. See  \cite{barnichruzz,celestialtorus} for recent discussions on the role of the  celestial surface topology.}   

\item $\sigma_{zz}$ is the shear and $T_{zz}$ is the so called Geroch tensor. 
The news tensor is defined  as \footnote{From this perspective, the role of the Geroch tensor is to substract the non-decaying part of the ``naive news'' $N^{\prime}_{ab}\, =\, \partial_{u}\sigma_{ab} + T_{ab}$, thus ensuring finiteness of various quantities such as the total radiated energy flux. Geroch's original derivation of the ``corrected'' news tensor \eqref{defnews} was motivated by the need to work with quantities that are covariant under conformal rescalings \cite{geroch}.} 
\end{itemize}
\be
N_{z z} = \partial_u \sigma_{z z} \label{defnews}
\ee
with
\be
N_{zz} \stackrel{u \to \pm \infty}{=} O(1/|u|^{2+\epsilon}) \label{fallnews}
\ee
as in \eqref{ourfalloffs}.

We now analyse the Geroch tensor $T_{ab}$. For simplicity, we focus on the chiral sector corresponding to $T_{zz}$ but it can be readily generalised to include the $T_{\zb\zb}$ components as well.

The $T_{zz} \neq 0$ sectors can be generated from the  $T_{zz}=0$ one by considering  spacetime diffeomorphisms generated by  finite superrotations $z \to \phi(z)$.  Following \cite{comperelong} (see also \cite{compere,cp,sraction}) one finds the resulting $T_{zz}$ is given by minus the Schwarzian derivative of $\phi$ wrt to $z$,
\be
T_{zz} = \frac{\phi'''(z)}{\phi'(z)}- \frac{3}{2}\left(\frac{\phi''(z)}{\phi'(z)}\right)^2 . \label{Titophi}
\ee
An important property of $T_{zz}$ which will be central to this paper is that it is holomorphic in the entire complex plane (or in general meromorphic with poles of order 2) 
\be
 \partial_{\zb}T_{zz} = 0. \label{holT}
 \ee
The presence of a non-trivial $T$ has non-trivial implications for the constraints at ${\cal I}^+$. In particular, the CK condition (i.e. the vanishing of the magnetic part of  $\sigma^\pm_{zz} = \lim_{u \to \pm \infty} \sigma_{zz}$) is modified in a sector with  $T_{zz} \neq 0$ \cite{cp}.  This modification can be understood as follows. 

In the presence of $T$, the sphere derivative $\partial_{z}$ is twisted to a  superrotation-covariant derivative $D_z$ (also referred to as Weyl-covariant derivative \cite{barnichruzz}) such that the CK constraints are  implemented by the conditions \cite{cp}, 
\be
\sigma^\pm_{zz}  =  -2 D^2_z \sigma^\pm = -2 \partial_z^2 \sigma^\pm + T_{zz} \sigma^\pm , \label{ckcond}
\ee
for some scalars $\sigma^\pm$. See   \autoref{softspacesec} for further details on the superrotation-covariant derivative.

We will call by $\Gamma$ the space of all possible fields $(\sigma_{ab}, T_{ab})$, with $ab=zz$ and $\zb \zb$, satisfying  conditions  \eqref{fallnews}, \eqref{holT}, \eqref{ckcond} (and their complex conjugated versions),
\be
\Gamma = \{ (\sigma_{ab}, T_{ab}) : \eqref{fallnews}, \eqref{holT}, \eqref{ckcond} \} . \label{intialdefgamma}
\ee
There are two functions on $\Gamma$ that will play a key role in our analysis. These are the leading and sub-leading soft modes of the news tensor:
\ba
\Nbzero_{zz}(z,\zb)  &:=&  \int_{-\infty}^\infty  N_{zz}(u,z,\zb) du ,  \label{leadingsoft} \\
 \Nbone_{zz}(z,\zb) & :=&  \int_{-\infty}^\infty  u N_{zz}(u,z,\zb) du .  \label{subleadingsoft}
\ea
Let us now discuss the extended BMS symmetries on $\Gamma$. The spacetime action of  super-translations and superrotations translates into 
 \be
\begin{array}{llllll}
\delta_f \sigma_{zz}   &= &  f \partial_u \sigma_{zz}- 2 D^2_z f , \quad & \delta_V \sigma_{zz}&  =&  (\L_V -\alpha   + \alpha u \partial_u) \sigma_{zz}, \\
\delta_f T_{zz} &=&0, \quad & \delta_V T_{zz} &=& \L_V T_{zz} - 2 \partial_z^2 \alpha. \label{gBMSaction}
\end{array}
\ee
where $\alpha=1/2 (\partial_z V^z + c.c)$. We remind the reader that $V^{a}$ is an entire vector field with polynomial coefficients.
We also note that contrary to earlier representations \cite{btaspects,stromSR,clpilot},  $\sigma_{zz}$ transforms homogeneously under superrotations as  the  inhomogeneous term is absorbed in $T_{zz}$. The supertranslation and superrotation charges are then given by \cite{compere,cp}
\ba
P_f  & = & \int d^2 z  \int du  N^{zz} \delta_f \sigma_{z z}   + c.c.  \\
J_V & = & \int d^2 z \left( \int  du N^{zz} \delta_V \sigma_{z z}  +  \Pit^{zz}  \delta_V T_{zz} \right) + c.c. ,
\ea
 where
\be
 \Pit_{zz} := 2 \Nbone_{zz} + \frac{1}{2}(\sigma^+ \sigma^+_{zz} - \sigma^- \sigma^-_{zz} ).  \label{defPizz}
\ee
The quadratic term in \eqref{defPizz} was introduced in \cite{cp} to ensure the consistency condition
\be
\delta_f J_V + \delta_V P_f =0 \label{consistencycond}
\ee
is satisfied.\footnote{This condition only requires the existence of Poisson brackets, but is otherwise insensitive to the presence or absence  of a 2-cocyle in the charge algebra. In particular, the BT charges satisfy \eqref{consistencycond}.}  Finally,  we can think of $\Gamma$ as a constrained phase space by treating $\Pit_{ab}$ as the momentum conjugate to $T_{ab}$, subjected to the constraint equation (\ref{defPizz}).

The symplectic form on $\Gamma$ implementing the supertranslation and superrotation charges is given by
\be
\Omega= \int d^2 z \left( \int  du \delta N^{zz} \wedge \delta \sigma_{z z}  + \delta \Pit^{zz} \wedge \delta T_{zz} \right) + c.c. \label{defOmega0}
\ee
which satisfies
\be
\delta P_f = \Omega(\delta,\delta_f) , \quad \delta J_V = \Omega(\delta,\delta_V) . \label{delQsOm}
\ee
We conclude  by noting that, in the present case where $\delta T_{zz}$ is holomorphic and $\delta T_{\zb \zb}$ anti-holomorphic, we can add to $ \Pit_{zz}$ a total $\partial_z$ derivative without affecting the expressions above. We will use this freedom to  trade $\Pit_{zz}$ for
\ba\label{june161}
\Pi_{zz}  & := &   \Pit_{zz} - \partial_z( \partial_z \sigma^+ \sigma^- - \partial_z \sigma^- \sigma^+)  \nonumber\\
& = &  2 \Nbone_{zz} + \frac{1}{2} (\sigma^+ + \sigma^-) \Nbzero_{zz} .
\ea
We end this section with three remarks.
\begin{itemize}
\item $\Pi_{zz}$ defined in Eq. (\ref{june161}) is a sum of leading and sub-leading soft news terms, such that  the coefficient of the leading soft news in the sum is  in fact the constant shear mode $C$. $\Pi_{zz}$ (and $\Pi_{\zb\zb}$) thus have a cleaner physical interpretation as compared to $\tilde{\Pi}_{zz}$. As we will  show this change from $\tilde{\Pi}_{zz}$ to $\Pi_{zz}$ will  play an important role in simplifying the analysis below, when we determine the reduced  phase space. 
\item We emphasise that this is one possible parametrisation of $\Pi_{zz}$ indicating one among many possible extensions of $\Gamma_{0}$. This is consistent with the well known  fact that consistency of Poisson brackets and requirement of faithful representation of asymptotic symmetries does not uniquely fix the radiative phase space \cite{stromST}. 
\item It is important to note that the homogenous change in $T_{zz}$ induced by infinitesimal holomorphic superrotations, $\delta_{V}\, T_{zz}\, +\, \partial_{z}^{3}\, V^{z}$, is holomorphic if and only if  the background $T_{zz}$ is holomorphic. Although generically $T_{ab}$ is meromorphic, in this paper we restrict ourselves to holomorphic $T_{zz}$ (or anti-holomorphic $T_{\zb\zb}$ ) tensors. 
\item To summarise, we started with the Ashtekar-Streubel radiative phase space  $\Gamma_{0}$ and extended it by  adding modes associated to superrotation symmetries, namely the pair $(\Pi_{ab}, T_{ab})$, where $\Pi_{ab}$ is \emph{constrained to be} a sum of leading and subleading soft news \eqref{june161}.  

\item We can thus think of $\Gamma$ as a subspace in a \emph{kinematical} phase space 
\be
\Gamma \subset \{\,  \sigma_{ab}(u, \hat{x}), \Pi_{ab}(\hat{x}),\, T_{ab}(\hat{x}) \}.
\ee
In the same spirit of the HLMS construction, we will use this description in order to obtain Poisson brackets via the Dirac procedure. 

\end{itemize}

 \section{Kinematical phase space}\label{sec4}
 
 
 The phase space $\Gamma$ introduced in the previous section still suffers from the  drawback in the AS phase space $\Gamma_0$ regarding the  Poisson bracket realization of supertranslations. In addition, 
 the mode conjugate to $T_{ab}$ has not been determined so far as $\Pi_{ab}$ is associated to the news tensor via a constraint. Following HLMS, our strategy to solve both issues will be to isolate the supertranslation and superrotation modes from the phase space $\Gamma$. This will lead us to a kinematical phase space $\Gkin$ with  well defined Poisson brackets that are consistent with eBMS transformations. The objective of the present section  is to introduce this kinematical phase space and brackets. In the next section we will follow the Dirac procedure to  obtain eBMS-compatible Poisson brackets on  $\Gamma \subset \Gkin$.

Our first step  will be to isolate the zero mode of the shear $\sigma_{ab}(u,\xh)$.\footnote{Our conventions differ from HLMS in two respects: (i) the zero mode is given by the average of the shear at $u = \pm \infty$ and (ii) a multiplicative  $(-2)$ factor in the definitions of  $C$ and $\Nzero$ \eqref{magconds}. The second choice  implies $\delta_f C= f$ under supertranslations.} 
We thus start by rewriting the shear as
\be
\sigma_{zz}(u,z,\zb) = \mathring{\sigma}_{zz}(u,z,\zb) +C_{zz}(z,\zb) \label{zeromode}
\ee
with  $\mathring{\sigma}_{zz}$ obeying
\be
 \mathring{\sigma}^+_{zz}+ \mathring{\sigma}^-_{zz}=0 \quad , \quad    \mathring{\sigma}^\pm_{zz} := \lim_{u \to \pm \infty} \mathring{\sigma}_{zz}(u,z,\zb). \label{sigmazerocond}
\ee
Note that the news \eqref{defnews} can be entirely written in terms of $\mathring{\sigma}_{zz}$,
\be
N_{z z} = \partial_u \mathring{\sigma}_{z z} . \label{defnews2}
\ee
A simple way to distinguish $\sigmao_{ab}$ from $\sigma_{ab}$ is that $\sigmao_{ab}$ contains soft news but no constant shear. The asymptotic condition \eqref{ckcond} can then be written as
\ba \label{magconds}
C_{zz}  =  -2 D^2_z\,  C\, ,  \quad \Nbzero_{zz}=  -2 D^2_z \Nzero  \label{CN}
\ea
for  scalars $C$ and $\Nzero$ such that
\be
C = \frac{1}{2}(\sigma^+ + \sigma^-), \quad \Nzero = {\sigma}^+ - {\sigma}^-.
\ee
In the parametrization \eqref{zeromode}, the symplectic form \eqref{defOmega0} now takes the form
\be
\Omega= \int d^2 z \left( \int  du \delta N^{zz} \wedge \delta \mathring{\sigma}_{z z}  + \delta \Nbzero^{zz} \wedge \delta C_{zz} + \delta \Pi^{zz} \wedge \delta T_{zz} \right) + c.c. , \label{defOmega}
\ee
with  $\Pi_{zz}  = 2 \Nbone_{zz} + C \Nbzero_{zz}$. 

The eBMS action on the shear sector becomes
\be
\begin{array}{llllll}
\delta_f \mathring{\sigma}_{zz}  &= &  f \partial_u \mathring{\sigma}_{zz} , \quad & \delta_V \mathring{\sigma}_{zz}&  =&  (\L_V -\alpha   + \alpha u \partial_u)\, \mathring{\sigma}_{zz}, \\
\delta_f C &=& f, \quad& \delta_V C &=& (\L_V - \alpha)\, C . \label{gBMSactionsigmazeroC}
\end{array}
\ee
We note that the inhomogeneous part of supertranslations is absorbed in $C$ so that $\mathring{\sigma}_{zz}$ transforms homogeneously under both supertranslations and superrotations. 



We will now use the splitting \eqref{zeromode} to define auxiliary phase spaces $\Gamma^\hard$ and $\Gamma^\soft$, each with its own symplectic structure and eBMS action. The kinematical phase space will be defined as $\Gkin = \Gamma^\hard \times \Gamma^\soft$.  The space $\Gamma$ will then arise as a constrained subspace on this kinematical space.


\subsection{Hard phase space} \label{hardspacesec}
We define $\Gamma^\hard$ as the space parametrized by $\mathring{\sigma}_{ab}$ 
\be\label{fallnews1}
\Gamma^\hard := \{ \mathring{\sigma}_{ab} :   N_{ab}(u, z,\zb) \equiv \partial_u \mathring{\sigma}_{zz}(u,z,\zb) \, \overset{u \to \pm \infty}{=}\, O(1/|u|^{2 + \epsilon})\, \} \ee
with symplectic structure
\be
\Omega^\hard  =  \int  d^2 z du \delta N^{zz} \wedge \delta \mathring{\sigma}_{zz} + c.c. 
\ee
At this stage we do not impose neither the zero mode condition $\mathring{\sigma}^+_{zz}+ \mathring{\sigma}^-_{zz}=0$ nor the CK condition. Both conditions will be imposed   in the next section  as part of the constraints that define $\Gamma$. Thus,  $\Gamma^\hard$ appears to be identical to the AS phase space $\Gamma_0$. It however differs in how  supertranslation are represented. Following \eqref{gBMSactionsigmazeroC}, we define the eBMS action on $\Gamma^\hard$  by
\be
\delta_f \mathring{\sigma}_{zz}  =   f \partial_u \mathring{\sigma}_{zz} , \quad  \delta_V \mathring{\sigma}_{zz}  =  (\L_V -\alpha   + \alpha u \partial_u) \mathring{\sigma}_{zz},
\ee
so that supertranslations (and superrotations) act homogenously on $\Gamma^\hard$. It is easy to verify  that this action is symplectic and generated by the charges
\ba
P^\hard_f & = & \int du d^2 z N^{zz}\delta_f \mathring{\sigma}_{zz} + c.c.  \\
  J^\hard_V  & = & \int du d^2 z N^{zz}\delta_V \mathring{\sigma}_{zz} + c.c. 
\ea

\subsection{Poisson brackets in the hard phase space} \label{hardPBsec}

To obtain Poisson brackets (PBs) involving the news, we start by considering its  Hamiltonian vector field (HVF),
\be
X_{N_{zz}(z,u)} = \frac{1}{2} \frac{\delta}{\delta \mathring{\sigma}_{\zb \zb}(u,z)}. \label{Xnews}
\ee
To simplify expressions, here and in the following we will use  $z$ rather than $(z,\zb)$ to denote points on the celestial sphere. It is well known that the shear does not admit a HVF  \cite{stromST,clgbms}. We can however define its PB with the news using \eqref{Xnews} leading to
\be
\{N_{zz}(z,u), \mathring{\sigma}_{\wb \wb}(u',w) \} = -\frac{1}{2} \delta(u-u') \delta^{(2)}(z,w),\label{newsshearpb}
\ee
and $\{N_{zz}(z,u), \mathring{\sigma}_{w w}(u',w) \}=0$. On the other hand, the non-zero PBs between the news tensor and itself is given by
\be
\{N_{zz}(z,u), N_{\wb \wb}(u',w) \} = \frac{1}{2} \delta'(u-u') \delta^{(2)}(z,w). \label{newsnewspb}
\ee

Other quantities of interest are the leading \eqref{leadingsoft} and subleading \eqref{subleadingsoft} soft modes $\Nbzero_{zz}$ and $\Nbone_{zz}$. Their HVFs are  given by\footnote{Readers not interested in the rigours of symplectic geometry of radiative phase space are encouraged to skip these subtleties and focus solely on the Poisson brackets between different modes.} 
\be
X_{\Nbzero_{zz}} = \int du  \frac{\delta}{\delta \mathring{\sigma}_{\zb \zb}(u,z)} , \quad
X_{\Nbone_{zz}}  =  \frac{1}{2} \int du\, u \frac{\delta}{\delta \mathring{\sigma}_{\zb \zb}(u,z)} \label{XNbone}
\ee
Strictly speaking, the second HVF is ill defined on $\Gamma^\hard$ as it does not preserve the fall-offs in \eqref{fallnews1}. We will use it cautiously to define various PBs, often requiring regularization prescriptions. We shall then verify that these choices lead to a self-consistent set of Poisson brackets on the physical phase space. 

We now discuss the  PBs with the soft modes. We  only focus on brackets involving tensor components of different helicities, since those with the same helicity  are trivially vanishing.  

The leading soft mode clearly Poisson commutes with the news and with itself,
\be
\{ N_{zz}(u,z) , \Nbzero_{\wb \wb} \}  =0, \quad \{ \Nbzero_{zz} , \Nbzero_{\wb \wb} \}  =0.
\ee
The bracket between the news and the subleading mode is found to be
\be
\{ N_{zz}(u,z) , \Nbone_{\wb \wb} \}  = \frac{1}{2} \delta^{(2)}(z,w).
\ee
For the brackets involving  a leading and a subleading mode, we take the prescription $\{ \Nbzero_{zz} , \Nbone_{\wb \wb} \} := - X_{\Nbzero_{zz}}(\Nbone_{\wb \wb})=0$.\footnote{Notice that $X_{\Nbone_{zz}}(\Nbzero_{\wb \wb})$ is formally proportional to $\delta^{(2)}(z,w) \int^\infty_{-\infty} du$. Our prescription may  be thought of as a regularization of this divergence.} For the brackets involving the subleading mode with itself, we notice that  $X_{\Nbone_{zz}}(\Nbone_{\wb \wb})$ is formally proportional to $\delta^{(2)}(z,w) \int^\infty_{-\infty} u du$. This can  naturally be interpreted as vanishing. Summarizing, our prescription for the subleading mode  is such that it Poisson commutes with the leading mode and with itself,
\be
\{ \Nbzero_{zz} , \Nbone_{\wb \wb} \}  = \{ \Nbone_{zz} , \Nbone_{\wb \wb} \} =0.
\ee

We conclude the section by bringing  attention to the fact that the HVF operation may not be continuous with respect to expressions involving  $u=\pm \infty$ limits. For example, one has
\be
 \int_{-\infty}^\infty du\, X_{N_{z z}(z,u)}  = \frac{1}{2} X_{\int_{-\infty}^\infty du\, N_{z z}(z,u)} .
\ee
However, as we will see in subsequent sections, this technicality has no effect on the final symplectic structure defined on $\Gamma$. 

\subsection{Soft phase space} \label{softspacesec}
We now define what we call the soft phase space which is co-ordinatized by soft modes and their conjugate fields,
\be
\Gamma^\soft := \{( C, \Nzero, \None_{ab}, T_{ab})  \},
\ee
where $C$ and $\Nzero$ capture, respectively, the zero mode of the shear and the leading soft news.  $\None_{ab}$ represents the sub-leading soft mode of the news. We note that the vanishing of super-translation magnetic charge implies that the leading soft news is parametrized by a single scalar mode on the celestial sphere. However as the superrotation magnetic and electric charge coincide \cite{subsub2}, $\overset{1}{N}_{ab}$ has two independent modes. 

At this stage we treat these fields as independent from $\Gamma^{\textrm{hard}}$.  Their relation with the fields in the hard sector  will be  imposed later in terms of  the constraints
\ba
 \Nbzero_{ab}  & = & - 2 D_{a}D_{b} \Nzero  ,  \label{cond1} \\
 \Nbone_{ab}  & =& \None_{ab}  . \label{cond2}
\ea
Similarly, we do not yet require $T_{zz}$ to be holomorphic (and $T_{\zb \zb}$ antiholomorphic); this condition will  also be part of the constraints to be imposed later.

The action of eBMS on $\Gamma^\soft$ is given by
\be
\begin{array}{llllll} \label{gBMSactionsoft}
\delta_f C &=& f, \quad& \delta_V C &=& (\L_V - \alpha) C \\
\delta_f \Nzero &=& 0, \quad & \delta_V \Nzero &=& (\L_V - \alpha) \Nzero \\
\delta_f \None_{zz} &=& 2 f D^2_z \Nzero , \quad &  \delta_V \None_{zz} &=& (\L_V  - 2\alpha) \None_{zz} \\
\delta_f T_{zz} &=&0, \quad & \delta_V T_{zz} &=& \L_V T_{zz} - 2 \partial_z^2 \alpha.
\end{array}
\ee
The expressions for $C$ and $T_{ab}$ were already described in  \eqref{gBMSactionsigmazeroC} and \eqref{gBMSaction}, and those for $\Nzero$ and  $\None_{ab}$ can be obtained from the action on the news \cite{cp}.

From the splitting of $\Omega$ in \eqref{defOmega} and the analgous splitting for the charges, we define
\be
\Omega^\soft =   \int d^2 z  (\delta \Nzero^{zz} \wedge \delta C_{zz} + \delta \Pi^{zz} \wedge \delta T_{zz}) + c.c. \label{Omegasoft}
\ee
\ba
P^\soft_f  &  = & \int d^2 z \Nzero^{zz} \delta_f C_{zz} + c.c.  \label{defPsoft} \\
J^\soft_V & = & \int d^2 z (\Nzero^{zz} \delta_V C_{zz} + \Pi^{zz} \delta_V T_{zz}) + c.c. \label{defJsoft}
\ea
where
\be
C_{zz} = -2 D^2_z C, \quad \Nzero_{zz} = -2 D^2_z \Nzero,
\ee
\be
\Pi_{zz} = 2 \None_{zz} + C \Nzero_{zz}  .
\ee
We shall see below that \eqref{defPsoft}, \eqref{defJsoft} generate the eBMS action \eqref{gBMSactionsoft} on $\Gamma^\soft$.

In order to facilitate computations, in particular to verify that the above symplectic form and charges lead to a closed eBMS algebra, it will be useful to isolate $C, \Nzero$ from $\None_{ab}, T_{cd}$. This will be achieved through the use of certain identities involving the superrotation-covariant derivative and the Geroch tensor that we now review.

The superrotation-covariant derivative  $D_a$  is constructed with the help of a ``potential''  $\psi$ for the Geroch tensor (also referred to as Liouville field \cite{compere}) satisfying\footnote{In the parametrization of \eqref{Titophi} one has $\psi(z,\zb) = - 1/2\ln (\phi'(z) \bar{\phi}'(\zb))$. There is change in notation with respect to \cite{cp}: There, $D_a$ denotes the  sphere covariant derivative and $\bar{D}_a$ the superrotation-covariant derivative. The formulas of \cite{cp} can be translated to the ones here by the substitutions $D_a \to \partial_a$, $\bar{D}_a \to D_a$ and $\R \to 0$, where $\R$ is the scalar curvature of the 2d metric.}
\be
T_{zz}  = 2 ((\partial_z \psi)^2+ \partial^2_z \psi).
\ee
$D_a$ acts as the ordinary derivative $\partial_a$ plus two contributions: (i) A multiplicative term of the form $k \partial_a \psi$ with $k$ dictated by the $\alpha$ factor in the action of $\delta_V$ (for $C$ and $\Nzero$, $k=-1$) and (ii)  Christoffel-like symbols $\Gpsi_{a b}^{c}$ that take into account the tensorial structure of the field being acted upon. 
In the present setting the only non-zero symbols are $\Gpsi_{zz}^{z}=-2 \partial_z \psi$ and $\Gpsi_{\zb\zb}^{\zb}=-2 \partial_{\zb} \psi$. For example,
\ba
D_z C & =& \partial_z C - \partial_z \psi C \label{DzC} \\
D^2_z C & = & \partial_z D_z C - \partial_z D_z C - \Gpsi_{zz}^{z} D_z C \\
& = &  \partial^2_z C - \frac{1}{2}T_{zz} C.
\ea
With these rules one can compute superrotation-covariant derivatives of arbitrary order. Of particular interest for our purposes is the 4-th order differential operator
\ba
\D C & := &  4 (D^2_z D^2_{\zb} +  D^2_{\zb} D^2_z) C  \label{defcalD} \\
& = &  (8  \partial^2_z \partial^2_{\zb} - 4 ( T_{\zb \zb} \partial_z^2  +T_{z z} \partial_{\zb}^2 )  + 2   T_{zz} T_{\zb\zb}) C.
\ea
Note that these operators have the same form when acting on $\Nzero$.  From these expressions, one can verify the identities, 
\ba
\delta( \D C) &= &  \D \delta C - 4 (D^2_z C \delta T_{\zb \zb} + c.c.) , \label{deltaDCid} \\
\delta( D_z^2 C) & =& D^2_z \delta C -\frac{1}{2} C \delta T_{zz}, \label{deltaD2zC}
\ea
and identical identities involving  $\Nzero$ instead of $C$.  

Using the above definitions and identities, one can rewrite the symplectic form \eqref{Omegasoft} in the following alternative forms
\ba
\Omega^\soft  &=&  \int d^2 z  \Big(  \delta ( \D \Nzero ) \wedge \delta C +2 (\delta(\None_{zz}- 2 C D^2_z \Nzero) \wedge \delta T_{\zb \zb}) + c.c.)   \Big)  \label{Osoft1}\\
&=& \int d^2 z  \Big(   \delta  \Nzero  \wedge \delta( \D C) + 2(  \delta \None_{zz} \wedge \delta T_{\zb\zb} + c.c.).  \Big) \label{Osoft2}
\ea
Similarly, the soft charges can be rewritten in various ways. For instance
\ba
P^\soft_f & =& \int d^2 z f \D \Nzero ,  \label{Psoft1} \\
J^\soft_V & = & \int d^2 z \Big( - \D C  \delta_V \Nzero +2( \None_{zz}  \delta_V T_{\zb \zb} + c.c.) \Big). \label{Jsoft2}
\ea
We end this section with a few remarks.
\begin{itemize}
\item $\Omega^\soft$ as written in Eq. (\ref{Osoft2}) makes it explicit that (prior to imposing the constraints),  ${\cal D}C$ is conjugate to $\Nzero$ and $T_{zz}$ is conjugate to $\None_{\zb\zb}$. This will simplify the constraint imposition enormously. 
\item The first form of $\Omega^\soft$ in Eq. \eqref{Osoft1} is specially suited for analyzing supertranslations since
\be
\delta_f (\None_{zz}- 2 C D^2_z \Nzero) =0.
\ee
Together with $\delta_f T_{zz}=0$, and using the form  \eqref{Psoft1} for $P^\soft_f$  it immediately follows that
\be
\delta P^\soft_f  = \Omega^\soft(\delta,\delta_f).
\ee
\item Finally, using \eqref{Osoft2} and \eqref{Jsoft2} it can be shown after a rather lengthy but straightforward computation that
\be
\delta J^\soft_V  = \Omega^\soft(\delta,\delta_V).
\ee
\end{itemize}
\subsection{Poisson brackets in the soft phase space}
In this section we analyse the Poisson brackets of ``elementary" fields in $\Gamma^{\textrm{soft}}$. 
We need  the Green's function $G(z,w)$ of the differential operator $\D$ defined in \eqref{defcalD} for this purpose,
\be
 \D G(z,w) = \delta^{(2)}(z,w). \label{defG}
\ee
Although we will not need the explicit form of $G(z,w)$, we know its leading term in the $T_{ab}$ expansion \cite{stromST}:
\be
 G(z,w)  = \frac{1}{16 \pi} |z-w|^2 \ln |z-w|^2 + O(T).
\ee
From the symplectic structure $\Omega^\soft$,  the HVFs of the elementary variables are found to be 
\ba
X_{C(z)}  & = & - \int d^2 w G(z,w)\Big(\frac{\delta}{\delta \Nzero(w)} + 2 D^2_w C(w) \frac{\delta}{\delta \None_{ww}(w)}+c.c. \Big)  , \label{XC} \\
 X_{\Nzero(z)} &= & \int d^2 w G(z,w)  \frac{\delta}{\delta C(w)},  \label{XNzero} \\
X_{\None_{zz}(z)} & = & \frac{1}{2} \frac{\delta}{\delta T_{\zb \zb}(z)}+  2 D^2_z C(z) \int d^2 w G(w,z) \frac{\delta}{\delta C(w)} ,  \label{XNone} \\
X_{T_{zz}(z)} & =& -\frac{1}{2} \frac{\delta}{\delta \None_{\zb \zb}(z)}  , \label{XT}
\ea
The HVFs for  $X_{\None_{\zb\zb}}$ and $X_{T_{\zb\zb}}$ can be written similarly.  We can now compute elementary Poisson brackets on $\Gamma^\soft$,
\ba
 \{C(z), \Nzero(w) \} &=& G(z,w), \\
 \{ C(z) , \None_{w w}(w) \}& =&  2 D^2_w C(w) G(z,w) , \\ 
   \{\None_{zz}(z), T_{\wb\wb}(w) \} &=& - \frac{1}{2}\delta^{(2)}(z,w),
\ea
together with their complex conjugated versions. All remaining brackets vanish.  This completes the characterization of the soft phase space.

\section{Poisson brackets on $\Gamma$}\label{secfour}
We now derive  the physical phase space at ${\cal I}^{+}$ which includes supertranslations as well as superrotation modes. Our strategy is the same as the one employed by HLMS \cite{stromST}, which is to impose certain constraints on the extended phase space which express the soft radiative modes as ``averages" of finite energy radiative data. However we now have more constraints thanks to the sub-leading soft sector parametrized by $(\, \None_{ab},\, T_{ab}\, )$.
For the benefit of the reader, we summarize the  $4 \times (2 \times \infty$) constraints once again: 
\begin{flalign}\label{june201}
\begin{array}{lll}
F^0_{ab}(z,\zb) \, :=\, \Nbzero_{ab}(z,\zb)\, +\, 2\, D_{a}D_{b} \Nzero(z, \zb)\\
F^1_{ab}(z,\zb)\, :=\, \Nbone_{ab}(z,\zb)\, -\, \None_{ab}(z,\zb)\\
F^2_{a}(z,\zb)\, :=\, \partial^{b}T_{ab}(z, \zb)\\
F^3_{ab}(z, \zb)\, :=\, \mathring{\sigma}^{+}_{ab}(z,\zb) + \mathring{\sigma}^{-}_{ab}(z,\zb)
\end{array}
\end{flalign}
where we recall that $ab = zz$ or $\zb \zb$. We regard \eqref{june201} as functions on the kinematical phase space 
\be
\Gkin = \Gamma^\hard \times \Gamma^\soft \quad , \quad \Okin=\Omega^\hard + \Omega^\soft.
\ee
By construction, the constraint surface $\{ F^I=0, I=0,1,2,3\}$ defines the embedding of the phase space $(\Gamma,\Omega)$ of section \ref{sec3} inside $\Gkin$, namely,
\be
\Gamma = \Gkin|_{\{F^I=0\}} \quad, \quad \Omega = \Okin|_{\{F^I=0\}}.
\ee
Since the Poisson brackets on $\Gkin$ are known, we can follow Dirac's procedure to obtain Poisson brackets on $\Gamma$. We do so in appendix \ref{appA}. There, we show that the constraints are indeed second class so that we can apply Dirac's formula. It is found that from all possible brackets among  elementary variables, only three get modified, namely
\begin{flalign}\label{dbrp}
\begin{array}{lll}
\{\, N_{\zb\zb}(u),\, \None_{ww}\, \}_{\star}\, =\, \-\, O_{\wb}\, \delta^{(2)} (z,w)\\\{\, C(w),\, \None_{zz}\, \}_{\star}\, =\, 2 (1\, -\, O_{z}) D_{z}^{2}C(z)\, G(w,z)\\
\{\, \None_{zz},\, T_{\wb\wb}\, \}_{\star}\, =\, -\, \frac{1}{2}(1 - O_{w}) \delta^{(2)}(z,w) ,
\end{array}
\end{flalign}
where $O_{w}\, =\, \partial_{w}\, \partial_{w}^{-2}\, \partial_{w}$ and $O_{\wb}\, =\, \partial_{w}\, \partial_{\wb}^{-2}\, \partial_{\wb}$.\\
Together with their complex conjugated versions. All remaining Dirac brackets coincide with the kinematical ones, as presented in section \ref{sec4}.


Operators such as $(1\, -\, O_{w}\, )$ ensure that the constraint $\partial_{w}\, T_{\wb\wb}\, =\, 0$ is   satisfied. 
$O_{w},\, O_{\wb}$ are symmetric operators and have the same kernel and co-kernel. 
 $O_{w}\, =\, \partial_{w}\partial_{w}^{-2}\, \partial_{w}$ is the operator whose kernel and co-kernel is the space of  anti-holomorphic functions on the complex plane, and such that it reduces to the identity on the space of smooth functions (modulo the anti-holomorphic kernel).

In particular, this implies that if $T_{\wb\wb}$ is smeared with a function which is entire anti-holomorphic, than its bracket with $\None_{zz}$ vanishes. 

For completeness we present the remaining non-zero Dirac brackets
\begin{flalign}\label{dbrp1}
\begin{array}{lll}
\{N_{zz}(z,u), N_{\wb \wb}(u',w) \}_\star = \frac{1}{2} \delta'(u-u') \delta^{(2)}(z,w), \\
 \{C(z), \Nzero(w) \}_\star = G(z,w), 
\end{array}
\end{flalign}
whose expression coincide with the kinematical ones. 

Eqs. (\ref{dbrp}), (\ref{dbrp1}) provide the non-vanishing Poisson brackets on $\Gamma$. They can be thought of as an extension of the HLMS brackets that incorporates the subleading soft modes $T_{ab}$ and $\None_{cd}$. We observe a crucial structural difference with the HLMS space: The hard modes $N_{ab}(u,z,\zb)$ do not Poisson commute with $\None_{ab}$, whereas  in $\Gamma_{\textrm{HLMS}}$  these hard degrees of freedom have vanishing Poisson brackets with the soft modes $C(z,\zb)$ and  $\Nzero(z,\zb)$. As we will see below, this has rather non-trivial consequences in the quantum theory. The existence of the sub-leading soft modes imply that the soft Hilbert space and finite energy graviton Fock space do not factorise.

.

\section{Quantization of $\Gamma$}\label{secfive}

In order to quantize the space $\Gamma$, we start by choosing a family of elementary phase space functions to be promoted to operators. We consider the Fourier transform of the news tensor
\be
N_{zz}(\w):= \int_{-\infty}^\infty du e^{i \w u}  N_{zz}(u,z,\zb),
\ee
together with the soft sector variables. That is, we seek a Hilbert space realization of operators $\Nh_{zz}(\w), \Ch(z), \widehat{N}^0(z),  \Th_{zz},  \Noneh_{zz}$ with commutators dictated by the Dirac brackets (\ref{dbrp}), (\ref{dbrp1}):
\be
[\Nh_{zz}(\w),\Nh_{\wb \wb}(\w')] =   \hbar \pi \, \w \delta(\w+\w') \delta^{(2)}(z,w) ,
\ee
\be
[\Nh_{zz}(\w),\Noneh_{\wb \wb}] = i \hbar \pi \delta(\w)\, O_{w}\, \delta^{(2)}(z,w) \label{newsN1comm}
\ee
\be
 [\Ch(z), \Nzeroh(w) ] = i \hbar \Gh(z,w) \label{CNzerocomm}
 \ee
 \be
[ \Ch(z) , \Noneh_{w w} ] =  2 i \hbar (1\, -\, O_{w}\, ) \Dh^2_w \Ch(w) \Gh(z,w) , \label{CNonecomm}
 \ee
 \be
[\Noneh_{zz}, \Th_{\wb\wb} ] = - \frac{i \hbar}{2} (1- O_{w}\, ) \delta^{(2)}(z,w), \label{N1Tcomm}
\ee
with the corresponding hermitian conjugated versions and with all remaining commutators being zero. Both the Green's function and the superrotation-covariant derivative become operators due to their dependence on the Geroch tensor. The RHS of \eqref{CNonecomm}, however, does not present operator order ambiguities since $\Ch$ and $\Th_{zz}$ commute.

We start by constructing a Hilbert space based on the following set of maximally commuting operators:
\be
\{\Nh_{zz}(\w) , \Nh_{\zb \zb}(\w),  \w>0 \} , \quad \widehat{N}^0(z), \quad   \Th_{zz}, \Th_{\zb \zb}
\ee
This choice is motivated by the fact that News tensors and the displacement memory are physical observables in the theory. We then consider a Hilbert space of the form
\be
\H = \H^\hard \otimes \H^\soft,
\ee
with $\H^\hard$ the graviton Fock space associated to the positive frequency news operators and $\H^\soft$ the space of wavefunctionals $\Psi(N,T)$ where $N \equiv \Nzero(z)$ and $T\equiv (T_{zz}, T_{\zb \zb})$ label the eigenvalues of the corresponding soft operators,
\ba
 \Nzeroh(z)\Psi(N,T) & = & N(z)\Psi(N,T) \\
 \Th_{zz}\Psi(N,T)   & = &  T_{zz}\Psi(N,T) \\
 \Th_{\zb\zb}\Psi(N,T)   & = &  T_{\zb\zb}\Psi(N,T).
 \ea 
Thus, a basis for $\H$ is given by states of the form
\be \label{basisH}
|(p_1,h_1), \ldots, (p_n,h_n) ; N , T \ket,
\ee
where the  $(p_i,h_i), i=1,\ldots,n$ label momenta and helicity of the Fock-space gravitons and $N,T$ are eigenvalues of the operatos $ \Nzeroh(z)$, $\Th_{zz}$ and $\Th_{\zb\zb}$. 

We will occasionally  find it convenient to work with states that are smeared over the soft variables, for which we shall use the notation 
\be
|\{p_i,h_i \} \ket_{\Psi(N,T)}:=  \int d N dT \Psi(N,T) |\{p_i,h_i \}; N , T \ket. \label{fullstates}
\ee
 These states can be interpreted as graviton Fock states ``dressed'' by  soft wavefunctionals $\Psi(N,T)$, similar to the fiber-bundle structure advocated in \cite{aalectures}.

We now describe the remaining operators. The zero-mode shear $\Ch(z)$ is naturally defined as an operator that is trivial on $\H^\hard$ and that on $\H^\soft$ acts according to 
\be
 \Ch(z) \Psi(N,T)  =  i \hbar \int d^2 w G_T(z,w) \frac{\delta}{\delta N(z)} \Psi(N,T), \label{defChat}
 \ee
 thus ensuring the commutator \eqref{CNzerocomm}. We use the notation $G_T$ for the Green's function in order to make explicit its dependence  on the Geroch tensor label $T$. The subleading news operator is more subtle, as it is sensitive to both soft and hard sectors. The commutators \eqref{newsN1comm} and \eqref{N1Tcomm} suggest the  definition
 \be
 \Noneh_{zz} |\{(p_i,h_i) \}  \ket_{\Psi(N,T)} = \big(-i O_{z}\, \lim_{\w \to 0} \partial_\w \Nh_{zz}(\w) - \frac{i\hbar}{2}(1- O_{z})\, \frac{\delta}{\delta T_{\zb \zb}}  \big)|\{(p_i,h_i) \}  \ket_{\Psi(N,T)}. \label{defNonehat}
 \ee
By construction, this expression satisfies \eqref{newsN1comm} and \eqref{N1Tcomm}.  One can further check that \eqref{defNonehat} and \eqref{defChat} satisfy \eqref{CNonecomm}. It is also easy to see that the above operators  reproduce the vanishing commutator relations dictated by the Dirac brackets. 
 
\subsection{eBMS quantum charges}
Promoting the classical expressions to operators we have
\ba
\Ph_f & =& \Ph^\hard_f+ \Ph^\soft_f \\
\Jh_V & =& \Jh^\hard_V+ \Jh^\soft_V
\ea
We define the hard charges as the standard Fock-space hard charges used in the literature (see e.g. \cite{stromST,stromSR,dfh}). These act non-trivially on the hard Hilbert space factor according to:
\ba\label{cssoft2}
\Ph^\hard_f |\{(p_i,h_i) \} ; N, T \ket & = & \sum_{i} E_i f(z_i,\zb_i) |\{ p_i,h_i \}  ; N, T \ket \\
\Jh^\hard_V |\{(p_i,h_i) \} ; N, T \ket & = & -i \hbar \sum_{i} \delta^i_V |\{ p_i,h_i \}  ; N, T\ket
\ea
where $(E_i,z_i,\zb_i)$ parametrize the momenta of the $i$-th graviton and $\delta^i_V = V^{z_i} \partial_{z_i} + V^{\zb_i} \partial_{\zb_i} \pm (\partial_{z_i} V^{z_i}-\partial_{\zb_i} V^{\zb_i}) - 1/2(\partial_{z_i} V^{z_i}-\partial_{\zb_i} V^{\zb_i}) E_i \partial_{E_i}$.  

For the soft charges we consider
\ba\label{june163}
\Ph^\soft_f & = &  \int d^2 z f(z) \widehat{\D} \Nzeroh(z),  \\
\Jh^\soft_V & = &  \int d^2 z \Big( - \delta_V \Nzeroh(z) \widehat{\D}\Ch(z)   +2( \delta_V \Th_{z z} \Noneh_{\zb\zb}   + h.c.) \Big).
\ea
$\Ph^\soft_f$ is purely multiplicative and so it does not present operator order ambiguities (the differential operator $\D$ becomes a multiplicative Hilbert-space operator $\Dcalh$ due to its dependence on the Geroch tensor).  For $\Jh^\soft_V$ we choose an operator ordering such that multiplicative operators are on the left and derivative operators on the right. This ordering choice will be justified a posteriori, as we will show that the eBMS algebra closes for $\Jh_{V}^{\textrm{soft}}$ defined as in \eqref{june163}.

\subsection{Ward identities of eBMS charges} 
Before proving that the quantized eBMS algebra closes, we first analyse  the Ward identity associated to $\Ph_{f},\, \Jh_{V}$ for tree-level S matrix. The equivalence of Ward identities for supertranslations and superrotations with leading and sub-leading soft graviton theorem has been extensively analysed in the literature, where the quantized soft charges are defined via soft limits of finite energy insertions, with the seminal references being \cite{stromST,stromSR}.  However, in the last few years, the quantization of $\Ph_{f}$ in the extended phase space $\Gamma_{\textrm{HLMS}}$ \cite{stromIR,akhoury1,akhoury2,acl,pate} has led to a novel implication of the supertranslation Ward identity. That the vacuum transition from incoming to outgoing scattering state in a generic gravitational scattering is non-trivial and is constrained by the infinity of conservation laws. These constraints are of course consistent with Weinberg soft theorem, but can be interpreted as a statement in a ``dual basis" in which the scattering states are eigen-states of soft news operator rather than being displaced by it (via a soft graviton insertion). 

We review the basic idea of \cite{stromIR,akhoury1,akhoury2} below and then show that the quantization of $\Jh_{V}$ on $\Gamma$ leads to an analogous implication for the superrotation Ward identity. 


For the purpose of this section, we adopt the following notation for basis states. We will denote the generic outgoing scattering state as $\vert\, \textrm{out},\, N , T\, \rangle$. Here $\textrm{out}$ denotes an outgoing multi-graviton state where gravitons have arbitrary helicity and \emph{hard} momenta,\footnote{Hard momenta graviton states simply corresponds to states created by $a_{\omega, \hat{p}}^{\dagger}$ where $\omega\, >\, 0$.} as in \eqref{basisH}. 
As discussed earlier, these states are eigenstates of hard graviton momentum operators as well as leading soft news and Geroch tensor.
We will also denote a state with trivial soft modes, namely $\vert\, \textrm{in}, \,  N=0,\, T=0\, \rangle$ as $\vert\, \textrm{in}\, \rangle$. In particular, we will consider all the in-coming states to have trivial soft modes.

The Ward identity associated to $\Ph_{f}$ can finally be written as,
\begin{flalign}\label{june162}
\langle\, \textrm{out},\, N,\, T\, \vert\, [\Ph_{f}^{\textrm{soft}},\, S\, ]\, \vert\, \textrm{in}\, \rangle\, =\, -\,  \langle\,  \textrm{out},\, N,\, T\, \vert\, [\Ph_{f}^{\textrm{hard}},\, S\, ]\, \vert\, \textrm{in}\, \rangle\, ,
\end{flalign}
where the soft charge is defined in Eq. (\ref{june163}). 
The LHS of the Ward identity in Eq. (\ref{june162}) can thus be evaluated as,
\begin{flalign}
\langle\, \textrm{out},\, N,\,  T\, \vert\, [\Ph_{f}^{\textrm{soft}},\, S\, ]\, \vert\, \textrm{in}\, \rangle\, =\,  \int\, d^{2}z\, {\cal D}f(z)\, N(z)\, \langle\, 
\textrm{out}, \, N,\, T\, \vert\,  S \, \vert\, \textrm{in}\, \rangle\,
\end{flalign}
Thus, the supertranslation Ward identity can be written as, 
\begin{flalign} \label{STWardId}
\int\, d^{2}z\, {\cal D}f(z)\, N(z)\, \langle   \textrm{out},\, N,\, T\, \vert\,  S\,  \vert\, \textrm{in}\, \rangle\, =\, -\,\langle \textrm{out}, \, N, \, T \, \vert\, [\Ph_{f}^{\textrm{hard}},\, S\, ]\, \vert\, \textrm{in}\, \rangle\, .
\end{flalign}
As was shown in \cite{stromIR,akhoury1,akhoury2}, this identity thus constrains $N$ in terms of the eigenvalue of the hard supertranslation charge which is effectively the Weinberg soft factor.\\

The Ward identity associated to $\Jh_{V}$ presents two challenges. Since the contribution to the soft charge that is linear in $\None_{ab}$ exponentiates to a ``translation operator" in the space of Geroch tensors,  the relationship of this Ward identity with Cachazo-Strominger soft theorem is not immediately clear.\footnote{This subtlety parallels the subtlety in relating the two versions of supertranslation Ward identities, namely the one which constrains the soft vacuum transitions (reviewed above) with  the one leading to Weinberg's soft theorem. That is, 
in the standard Fock-space formulation of the Ward identity, one defines the superrotation soft charge in terms of $\lim_{\omega\, \rightarrow\, 0}\partial_{\omega}\, \w \langle\, \textrm{out}\, \vert\, a_{\omega, \hat{p}}\, S\, \vert\, \textrm{in}\, \rangle$. But in the new quantization of $\Jh_{V}^{\textrm{soft}}$, the term linear in $\Noneh_{ab}$ is defined as a sum of  the zero frequency operator and $\frac{\delta}{\delta T_{zz}}$ which may be interpreted as $\langle\, \textrm{out}\, \vert\, \lim_{\omega\, \rightarrow\, 0}\partial_{\omega}\, \omega\, a_{\omega, \hat{p}}\, S\, \vert\, \textrm{in}\, \rangle$. Depending on the poles of $V$, only one of the two terms contribute to the soft charge.} Furthermore,   $\Jh_{V}^{\textrm{soft}}$ contains quadratic operators involving the leading soft modes and it is not a priori  clear whether the corresponding Ward identity is affected by the inclusion of these new terms.

In fact, the superrotation soft charge appears to mix the hard and soft Hilbert space factors due to the subleading soft news operator \eqref{defNonehat}. However, when one studies the action of $\Jh^\soft_V$ \eqref{defJsoft} on a state \eqref{fullstates}, one finds only one of the two terms in $\hat{N}^{1}_{zz}$ is non-trivial.
\begin{itemize}
\item If $\delta_{V}T$ is holomorphic, then one finds the hard-sector part of $\Noneh_{zz}$ drops out due to the holomorphicity of $\delta_V T_{zz}$. The action is then found to be given by
\begin{flalign}
\begin{array}{lll}
 \Jh^\soft_V |\{(p_i,h_i) \} \ket_{\Psi(N,T)}=\\
\hspace*{0.5in}- i \hbar \int d^2 z \Big( \delta_V N(z) \frac{\delta}{\delta N(z)}   +( \delta_V T_{z z} \frac{\delta}{\delta T_{z z}}  + c.c.) \Big) |\{(p_i,h_i) \}  \ket_{\Psi(N,T)}, \label{Jsoftaction}
\end{array}
\end{flalign}
where we recall that $\delta_V N = (\L_V - \alpha) N$ and $\delta_V T_{zz}=\L_V T_{zz} - 2 \partial_z^2 \alpha$.
\item Even though $\hat{J}_{V}$ is strictly speaking only defined for holomorphic $\delta T_{ab}$, we now consider an extension of its domain of validity to meromorphic vector fields.\footnote{The reason we expect our charges to give the correct Ward identity for all $V^{a}$ is simply that we do not expect the subtlety of the change in the sphere metric to affect the single Ward identity associated to infinitesimal superrotations. The added subtlety would show up when we consider `higher order" structures such as commutators of two charges.}
 It can be immediately be verfied that if $\delta_{V}T$ has  poles at finite $z$, then the second term in $\hat{N}^{1}_{zz}$ drops out and the soft charge can be written as,
\begin{flalign}
\begin{array}{lll}
 \Jh^\soft_V |\{(p_i,h_i) \} \ket_{\Psi(N,T)}=\\
\hspace*{0.2in}- i \hbar \int d^2 z \Big( \delta_V N(z) \frac{\delta}{\delta N(z)}   -\,  O_{z}\, \delta_{V} T_{zz} \lim_{\w \to 0} \partial_\w \Nh_{zz}(\w)\, + c.c.) \Big) |\{(p_i,h_i) \}  \ket_{\Psi(N,T)} \label{Jsoftaction2}
\end{array}
\end{flalign}
\end{itemize}
In the second case, the Ward identity of $\hat{J}_{V}$ implies the Cachazo-Strominger soft theorem. Hence in the rest of the section, we analyse the first case in which the soft superrotation charge has a non-trivial action on $T_{ab}$.
We now evaluate the Ward identity between out-states of the form $\vert\, \textrm{out},\, N,\, T\, =\, 0\, \rangle$ and in-states $\vert\, \textrm{in}\, \rangle$ as before. The computation presented here can be easily generalised to generic scattering states.  The action of the soft superrotation charge on the in and the out states is given by
\begin{flalign}\label{cssoft1}
\begin{array}{lll}
\Jh^\soft_V |\textrm{in} \, \rangle = - i \hbar \int d^2 z ( \delta_V T_{z z} \frac{\delta}{\delta T_{z z}}  + c.c.) \vert\, \textrm{in},\, N=\, 0\, T = 0\, \rangle\\
\Jh^\soft_V |\textrm{out},\, N,\, T=0\, \rangle = - i \hbar \int d^2 z ( \delta_V T_{z z} \frac{\delta}{\delta T_{z z}}  + c.c.) \vert\, \textrm{out},\, N,\, T = 0\, \rangle\\
\hspace*{2.5in} -i\, \hbar\,\int\, \delta_{V}N(z) \frac{\delta}{\delta N(z)}\, \vert\, \textrm{out},\, N,\, T\, =\, 0\rangle
\end{array}
\end{flalign}
As $\delta_{V}\, N(z)\, =\, (L_{V}\, -\, \alpha\, )\, N(z)$, the action of $\int\, \delta_{V}N(z)\, \frac{\delta}{\delta N(z)}$ on the $\vert\, \textrm{in}\, \rangle$ state vanishes. 

In fact we note that the state obtained by action of $\, \frac{\delta}{\delta N(z)}$ on the out state will not contribute to the Ward identity since $N$ is already constrained by the leading soft theorem. More in detail as, 
\begin{flalign}\label{june165}
\begin{array}{lll}
 \langle\, \textrm{out},\, N + \lambda\, \delta_{V}N, T = 0\, \vert\, S\, \vert\, \textrm{in}, \lambda\, \delta_{V}\, \rangle\, =\, 0\, \forall\, \lambda\\
 \implies\, \lim_{\lambda\, \rightarrow\, 0}\, \frac{d}{d\lambda}\, \langle\, \textrm{out},\, N + \lambda\, N, T = 0\, \vert\, S\, \vert\, \textrm{in}\, \rangle\, =\, 0
 \end{array}
 \end{flalign}
By combining Eqs. (\ref{cssoft1}), (\ref{cssoft2}) and (\ref{june165}) we get,
\begin{flalign}\label{june171}
\int\, d^{2}z\, \delta_{V}T_{zz}\, \langle\, \textrm{out}\, N, T=0 \vert\, [\, \frac{\delta}{\delta T_{zz}},\, S\, ]\, \vert\, \textrm{in}, N=0, T=0\, \rangle&=\\
&\hspace*{-1.3in}-\, \langle\, \textrm{out},\, N,\, T\, =\, 0\, \vert\, [\, \Jh_{V}^{\textrm{hard}},\, S\, ]\, \vert\, \textrm{in}\, \rangle 
\end{flalign} 
The LHS of (\ref{june171}) can be understood as 
\begin{flalign}
\lim_{\lambda\, \rightarrow\, 0}\, \frac{d}{d\lambda}\, [\, \langle\, \textrm{out}\, N, \lambda\, \delta_{V}T \vert\, S\, \vert\, \textrm{in}\, \rangle\, -\, \langle\, \textrm{out}\, N, T = 0 \vert\, S\, \vert\, \textrm{in},\, N = 0, \lambda\, \delta_{V}T \rangle\,\, ]
\end{flalign}
The Ward identity thus implies that the infinitesimal change $\delta_{V}T$ in the Geroch tensor is equivalent to the action of sub-leading soft graviton operator on hard scattering states.

As we saw above, Cachazo-Strominger soft graviton theorem is realised as a Ward identity of the superrotation charge in the case $\delta_{V}T$ is meromorphic. But if $\delta_{V}T$ is holomorphic (except pole at infinity) then the 
relationship between the superrotation Ward identity and the Cachazo-Strominger soft theorem deserves further scrutiny. 

We end this section with some  comments.
\begin{itemize}
\item The operator action of $\Noneh_{zz}$ is a sum of two operators which act respectively on the hard and  soft Hilbert space factors \eqref{defNonehat}. One is the ``familiar" sub-leading soft graviton insertion of the type $\lim_{\omega\, \rightarrow\, 0}\partial_{\omega}\,  \Nh_{zz}(\omega, \hat{x})$ and the other one is the soft operator which is linear in $\frac{\delta}{\delta T_{zz}}$. We see that for the case rigorously analysed in this paper, $\Noneh_{zz}$ acts as the latter but our preliminary analysis suggests that for a generic superrotation, it is precisely one of these two operators that contribute to the soft superrotation charge. This may offer a new perspective on the relationship between sub-leading soft graviton theorem and Ward identities, akin to the relationship between dressed vacua, Weinberg soft theorem and supertranslation symmetries. We leave a more thorough analysis of this aspect and its impact on the dressed states for future work. 

\item The second class constraint $F^1_{ab}\, =\, \Nbone_{ab}\, -\, \None_{ab}\, =\, 0$ was strongly imposed classically by using Dirac brackets. On the other hand, had we first quantized the extended phase space and imposed $F^1_{ab}$ at quantum level, it would imply (at least formally) that we  identify $\Noneh_{ab}$ with $\widehat{{\cal N}}^{1}_{ab}\, =  -i \lim_{\w \to 0} \partial_\w \Nh_{zz}(\w)$. This could bring the Ward identity in the more ``traditional form" of a sub-leading soft graviton theorem for all superrotations.
\item We chose the scattering states to be eigenstates of $\Nzeroh$ and  $\Th_{ab}$ since $\Noneh_{ab}$ does not commute with the news tensor $\Nh_{ab}(\omega, \hat{x})$. However this commutator has support only at $\omega = 0$ and hence it may be possible to work with eigenstates of $\Nzeroh$ and $\Noneh_{ab}$ instead. In such case, the superrotation Ward identity would  take a form analogous to \eqref{STWardId}, equating a vacuum transition parametrized by the difference of $N^{1}_{ab}(\xh)$ between in and out states with an  amplitude with no vacuum transition but with hard states ``super-rotated" by the angular momentum operators. 
\end{itemize}
\subsection{eBMS quantum algebra}

We now evaluate the commutator of the quantum eBMS charges. We start by noticing that the hard part of the operators satisfy the extended BMS algebra (see e.g. \cite{dfh}):
\ba
& [\Ph_f^\hard,\Ph_{f'}^\hard] & = 0  \\
& [\Ph_f^\hard,\Jh_V^\hard ]  & = - i \hbar \Ph^\hard_{V(f)} \\
& [\Jh_V^\hard,\Jh_{V'}^\hard] & =  i \hbar \Jh^\hard_{[V,V']} .
\ea
where $V(f)= (\L_V - \alpha) f$.

We now discuss the commutators involving the soft part the operators. The only missing operator action to complete this computation is that of  $\Ph_{f}^{\textrm{soft}}$. The soft supermomentum has a simple multiplicative action that is only sensitive to the soft Hilbert space,
\be
\Ph^\soft_f |\{(p_i,h_i) \}  \ket_{\Psi(N,T)} = P^\soft_f(N,T) |\{(p_i,h_i) \}  \ket_{\Psi(N,T)} \label{Psoftaction}
\ee
where 
\be \label{PsoftfNT}
P^\soft_f(N,T)=\int d^2 x f \D_T N
\ee
is the value of the classical soft supermomentum associated with leading news $N$ and Geroch tensor $T$. We use the notation $\D_T$ to make explicit the $T$-dependence of the differential operator $\D$ \eqref{defcalD}.

Eqs. (\ref{Psoftaction}), (\ref{Jsoftaction}) imply that the hard and soft charges commute with each other.\footnote{Notice this is technically different from the individual terms that contribute to the commutator in \cite{dfh}. 
This is related to the fact that our definition of soft superrotation charge includes a  contribution from the zero mode of the shear that is usually considered as part of the hard charge.}  To complete the computation of the charge algebra we need to evaluate the commutator of the soft charges. Since $\Ph^\soft_f$ acts multiplicatively  it follows that 
\be
[\Ph_f^\soft,\Ph_{f'}^\soft]=0.
\ee
Given the expression  \eqref{Jsoftaction} for the action of $\Jh^\soft$ it easy to show that
\be
[\Jh_V^\soft,\Jh_{V'}^\soft]  =  i \hbar \Jh^\soft_{[V,V']}
\ee
and
\be
[\Ph^\soft_f , \Jh^\soft_V] \Psi[N,T] = i \hbar \delta_V (P^\soft_f (N,T)) \Psi[N,T],
\ee
where $\delta_V$ acts on the $N$ and $T$ labels of $P_f (N,T)$. Thus, to prove 
the algebra closure  we need to show that
\be \label{srcovariancePf}
\delta_V P_f (N,T) = -P_{V(f)}(N,T). 
\ee
This identity follows in fact due to the superrotation-covariance of the classical charge \cite{cp}:
\be
\delta_V P_f (N,T)  =  \delta_V \int d^2 x f \D_T N =  \int d^2 x f  \delta_V( \D_T N)  = - \int d^2 x V(f)  \D_T N, \label{delVPfNT}
\ee
where 
\be
\delta_V (\D_T N) = (\L_V + 3 \alpha) \D_T N. \label{delVDTN}
\ee
  Notice that the $T$-dependence of the supermomentum, appearing through the use of the superrotation-covariant differential operator $\D_T$, is what ensures the transformation property  \eqref{delVDTN} that underlies \eqref{srcovariancePf}.  It is interesting to compare with the situation where one works in the $T=0$ sector and does not include $\delta_V T$ terms. Denoting such action by $\delta'_V$, one finds
\ba
\delta'_V P_f (N,T=0)  &:= & \int d^2 x f   \D_0 \delta_V N \\ 
&= & \int d^2 x f  \delta_V( \D_T  N )|_{T=0} - \int d^2 x f  \delta_V( \D_T)  N |_{T=0} \\
&= & -P_{V(f)}(N,T=0) - \K_{(f,V)}(N,T=0)
\ea
where we used \eqref{delVPfNT} and defined 
\be
 \K_{(f,V)}(N,T) :=  \int d^2 x f  \delta_V( \D_T)  N .
\ee
From   \eqref{deltaDCid} and \eqref{gBMSaction} one finds the non-closure term is given by
\be
 \K_{(f,V)}(N,T)=  - 4 \int d^2 x f ( \L_V T_{\zb\zb} -  \partial_{\zb}^3 V^{\zb}) D^2_z N  + c.c.   . \label{defcalK}
\ee
By setting $T=0$, one obtains a 2-cocycle term of the type found in \cite{btchargealgebra}.\footnote{The relation of \eqref{defcalK} with the BT cocycle is $\K_{(f,V)} = -4 K_{(f,V)}$. The different multiplicative factor is due to the extra dependence on the Geroch tensor when using $\Nzero_{ab}=-2 D_a D_b \Nzero$. The analogue of \eqref{defcalK} for the  BT case is  $K_{(f,V)} = -2\int d^2 x f \delta_V( D^a D^b )\Nzero_{ab}|_{T=0}$.}

\subsection{Consistency with consecutive double soft graviton theorems}\label{dfhsec}
In \cite{dfh}, Distler, Flauger and Horn gave an explicit proof of the non-closure of eBMS algebra in quantum theory. Their main result can be summarized as follows.
Given the consecutive double soft graviton theorem in tree-level scattering amplitudes at sub-leading (and sub-sub leading) order, one can compute the commutator of two such soft limits.  The resulting factorisation theorem is equivalent to the algebra generated by the (quantized) supertranslation and superrotation charges. 

More in detail, the authors in \cite{dfh} considered the following commutators obtained from double soft theorems. 
\begin{flalign}\label{dsc}
\begin{array}{lll}
[\lim_{\omega_{1}\, \rightarrow\, 0}\, \omega_{1},\, \lim_{\omega_{2}\, \rightarrow\, 0}\, \partial_{\omega_{2}}\, \omega_{2}\, ]\, 
\langle\textrm{out}, k_{1}, k_{2}\, \vert\, S\, \vert\, \textrm{in}\, \rangle\, =\, 
S(k_{1}, k_{2})\, \langle\textrm{out}\, \vert\, S\, \vert\, \textrm{in}\, \rangle\\
\vspace*{0.1in}
[ \lim_{\omega_{1}\, \rightarrow\, 0}\, \partial_{\omega_{1}}\, \omega_{1},\, \lim_{\omega_{2}\, \rightarrow\, 0}\, \partial_{\omega_{2}}\, \omega_{2}\, ]\, 
\langle\textrm{out}, k_{1}, k_{2}\, \vert\, S\, \vert\, \textrm{in}\, \rangle\, =\, 
S^{\prime}(k_{1}, k_{2})\, \langle\textrm{out}\, \vert\, S\, \vert\, \textrm{in}\, \rangle
\end{array}
\end{flalign}
Here $k_{i}\, =\, (\omega_{i}, \hat{k}_{i})$ and we have supressed the helicity information of all the gravitons for the sake of brevity.  The two commutators are associated to consecutive double soft theorems at sub-leading and sub-subleading orders respectively. 

The authors then analysed $S(k_{1}, k_{2}),\, S^{\prime}(k_{1}, k_{2})$ in complete detail for all possible helicity configurations of the soft gravitons and showed that it consists of terms which have\\ 
(1) ``soft" poles consistent with single soft theorem or\\
 (2) colinear poles arising when $k_{1},\, k_{2}$ become colinear.  
 
 The RHS of (\ref{dsc}) was shown to equal  the following matrix elements
\begin{flalign}\label{fdh3}
\begin{array}{lll}
S(k_{1}, k_{2})\, \langle\textrm{out}\, \vert\, S\, \vert\, \textrm{in}\, \rangle\, =\, \langle\textrm{out}\, \vert\, [\, [\Ph_{f},\, \Qh_{V}],\, S\,]\, \vert\, \textrm{in}\, \rangle\\
\vspace*{0.1in}
S^{\prime}(k_{1}, k_{2})\, \langle\textrm{out}\, \vert\, S\, \vert\, \textrm{in}\, \rangle\, =\, \langle\textrm{out}\, \vert\, [\, [ \Qh_{V_{1}},\, \Qh_{V_{2}} ],\, S\,]\, \vert\, \textrm{in}\, \rangle
\end{array}
\end{flalign}
for specific choices of parameters $f,V,V_1$ and $V_2$ that are dictated by the helicity and direction of the soft gravitons. 
In the above equations  $\textrm{out/in}$ are outgoing or incoming graviton states in the ``trivial sector" defined by $N\, =\, T_{ab}\, =\, 0$. 
We have denoted the superrotation charge as $\hat{Q}_{V}$ instead of $\hat{J}_{V}$ in Eq. (\ref{fdh3}). This is because in the $\Nzero\, =\, T\, =\, 0$ sector, $\hat{Q}_{V}^{\textrm{soft}}$ differs from $\hat{J}_{V}^{\textrm{soft}}$ by
\begin{flalign}\label{oldso}
\hat{Q}_{V}^{\textrm{soft}}\, =\, \hat{J}_{V}^{\textrm{soft}}\, +\, i \hbar \int d^2 z\, \delta_V N(z) \frac{\delta}{\delta N(z)} 
\end{flalign}
Eq. (\ref{fdh3}) is already a rather surprising result. As the LHS is simply the soft factor multiplied by the scattering amplitude and the RHS is the Ward identities associated to commutators of the symmetry generators. This is structurally different than the identification between a single soft theorem and Ward identities.
\footnote{We have written the equality in (\ref{fdh3}) slightly differently than in \cite{dfh}. This is simply for the sake of pedagogy and does not affect the final result which is our primary concern. The interested reader is encouraged to consult the original reference for more details.} 
In fact in \cite{dfh}, the striking nature of the above result was made more manifest as it was shown that this identity implies  non-closure of eBMS algebra as, 
\begin{flalign}
\begin{array}{lll}
\langle\textrm{out}\, \vert\, [\, [\Ph_{f},\, \Qh_{V}],\, S\,]\, \vert\, \textrm{in}\, \rangle\,
=\, - i \hbar \langle\, \textrm{in}\, \vert\, [\, \hat{P}_{V(f)}\, +\, \hat{K}_{(f, V)},\, S\, ]\, \vert\, \textrm{out}\, \rangle\\
\langle\textrm{out}\, \vert\, [\, [\Qh_{V_{1}},\, \Qh_{V_{2}}],\, S\,]\, \vert\, \textrm{in}\, \rangle\,
=\, i \hbar \langle\, \textrm{in}\, \vert\, [\, \hat{Q}_{[V_{1},V_{2}]}\, ,\, S\, ]\, \vert\, \textrm{out}\, \rangle\\
\end{array}
\end{flalign}
The analysis by Distler, Flauger and Horn was for the entire eBMS algebra including holomorphic superrotations.  Hence we can compare the result stated above with our analysis.

Using the modified superrotation charge derived in this paper the above identities can be rewritten as,
\begin{flalign}
\begin{array}{lll}
[\, \hat{P}_{f},\, \hat{J}_{V}\, ]\, =\,- i \hbar \hat{P}_{V(f)}\\
\vspace*{0.1in}
[\, \hat{J}_{V_{1}},\, \hat{J}_{V_{2}}\, ]\, =\, i \hbar \hat{J}_{[V_{1},V_{2}]}
\end{array}
\end{flalign}
We thus observe that first equation in (\ref{fdh3}) implies that
\begin{flalign}
S(k_{1}, k_{2})\, \langle\textrm{out}\, \vert\, S\, \vert\, \textrm{in}\, \rangle\, =\, \langle\textrm{out}\, \vert\, [\, [\Ph_{f},\, \Jh_{V}]\, +\, \hat{K}(f,V),\, S\,]\, \vert\, \textrm{in}\, \rangle\, 
\end{flalign}
With the new superrotation charge which generates closed eBMS algebra (albeit only for entire vector fields), \emph{the right hand side} of (the commutator) of sub-leading double soft theorem can be realised as matrix element of commutator of S matrix with $\, [\, \hat{P}_{f},\, \hat{J}_{V}\, ]\, +\, \hat{K}(f,V)$.
\section{Conclusions and outlook} \label{sec7}
Our main goal in this paper has been to analyse the quantum charge algebra associated to extended BMS symmetries in light of recent progress in the understanding of superrotation charges in classical gravity. In this section, we briefly summarize our results and highlight some of the open issues which have emerged out of our analysis and require further investigation.
\subsection{Summary of results}\label{secsix}
 Throughout this paper, we considered the eBMS algebra generated by supertranslations and holomorphic (anti-holomorphic) superrotations.  Although we expect our analysis to generalise to the complete eBMS and gBMS  algebra (for which the superrotations are generated by meromorphic or smooth vector fields respectively) we did not attempt to do so in this paper and offer a few comments in  \ref{tgbmsa} below regarding the possible complications we are likely to encounter in attempting such an extension. 
  
The  main ideas in this paper can be summarized as follows: Based on the analysis in \cite{cp}, we considered an extension of the radiative phase space of shear modes which contains the leading and sub-leading soft news and their conjugate partners. This extended phase space is not physical in the sense that the new variables are constrained in terms of the shear and news fields.\footnote{These constraints assumed an invariant complex structure on the celestial sphere which is only preserved by meromorphic vector fields.}\\
 We showed that these constraints are  second class and can be solved explicitly to obtain a physical radiative phase space $\Gamma$ on which both the supertranslation as well as superrotation charges have a well defined action.\\
 We then quantized the phase space and showed that the quantized charges form a closed algebra which is simply a  faithful (as opposed to projective) representation of the eBMS Lie algebra at ${\cal I}$.\\
 A natural off-shoot of our analysis is an explicit parametrization of the space of all vacua which generates (via finite energy excitations) a  representation space for the eBMS algebra. These vacua are labelled by the soft news $\Nzero$ and the Geroch tensor $T_{ab}$ which is a Schwarzian.  This result should thus be viewed as a synthesis  of the structure of vacua analysed in \cite{stromST} and \cite{sraction}. 

\subsection{Comparison with the eBMS algebra in celestial holography}\label{ebmscel}
We now offer some brief and rather speculative comments on the implication of our results for the eBMS algebra in celestial holography which has been derived and analysed in a number of recent papers (see \cite{ebmsope,confsoft1, confsoft2, confsoft3, shamik, alfredo})\footnote{We thank Dileep Jatkar for extensive discussions  on eBMS algebra in celestial CFT.}.  Our comments are rather preliminary and only serve to highlight a possible way in which  the quantized eBMS algebra at null infinity in this paper can be reconciled with the eBMS algebra in celestial CFT. 

To be more specific, we consider only  positive helicity soft insertions. The  holomorphic supertranslation current and the boundary stress tensor are defined as, 
\begin{flalign}
\begin{array}{lll}
P(z)\, =\, P(f)\vert_{f\, =\, \frac{z-w}{\zb - \overline{w}}}\\
{\cal T}_{\zb\zb}\, =\, Q_{Y}^{\textrm{soft}}\vert_{Y^{\zb}\, =\, \frac{1}{\zb-\overline{w}}, Y^{z}\, =\, 0}
\end{array}
\end{flalign}
We have denoted the stress tensor as ${\cal T}_{ab}$ to distinguish it from $T_{ab}$ which in this paper denotes  the conjugate partner to the sub-leading soft news.\\
In the celestial basis, $P(z)$ is a conformal primary with $\triangle = 1$ and ${\cal T}_{\zb\zb}$ is the shadow transform of a $\triangle = 0$ primary operator. \\
In the above equation, $w$ is being integrated over. We also emphasize that the soft charge $Q_{Y}^{\textrm{soft}}$ used to define the celestial stress tensor is the ``old" soft superrotation charge given in Eq. (\ref{oldso}). However in order to make contact with the results derived in this paper, we need to consider the following tensor on the celestial sphere 
\begin{flalign}
t_{\zb\zb}\, =\, J_{Y}^{\textrm{soft}}\vert_{Y^{\zb}\, =\, \frac{1}{\zb- \overline{w}}, Y^{z}\, =\, 0}
\end{flalign}
Once again, we have assumed that  our definition of $J_{Y}$ is valid for a meromorphic vector field, as we believe that  this has no bearing on the main point raised in this section.
With this tensor, we now know that the commutator between $[P(z),\, t_{\zb\zb}\, ]$ closes in a super-translation.\footnote{This result can be re-written  in the so-called celestial basis which is obtained by Mellin transforming $P_{z},\, t_{zz}$ to conformal operators.}\\
But as, 
\begin{flalign}
t_{\zb\zb}\, =\, {\cal T}_{\zb\zb}\, +\, {\cal O}_{\zb\zb}
\end{flalign}
The algebra generated by $P(z), {\cal T}_{\zb\zb}$ does not close and has a 2 co-cycle extension consistent with \cite{dfh}. 

We expect that in the celestial basis, the OPE of $t_{ab}$ with conformal primaries (with $\triangle\, =\, 1 + i\lambda,\, \lambda\, \neq\, 0$) is the same as that of ${\cal T}_{ab}$. This is because this OPE follows from sub-leading soft graviton theorem which as a Ward identity is insensitive to the distinction between the old and new superrotation charges. In other words, we expect that the following OPE has no singular terms.
\begin{flalign}
{\cal O}_{ab}\, O_{\triangle, \pm 2}(z,\zb)\, =\, \textrm{regular}
\end{flalign}
where $O_{\triangle, \pm 2}$ is the hard graviton in the celestial basis with conformal dimension $\triangle\, =\, 1 + i\lambda$. Thus both ${\cal T}_{ab},\, t_{ab}$ appear to be possible candidates for a  stress tensor on the celestial sphere. 

However as the algebra of $t_{ab}$ with $P(z)$ is isomorphic to the eBMS algebra in \cite{ebmsope}, we also expect that the OPE of $t_{ab}$ with the supertranslation current is precisely as expected of an OPE between a CFT stress tensor and a Kac-Moody current. That is, with $t_{ab}$ as the celestial stress tensor,  $P(z)$ is a level-1 descendent primary field. But we note that in the celestial basis  ${\cal O}_{ab}$ is the Mellin transform of,
\begin{flalign}
\int\, -\, {\cal D}C\, \delta_{Y}\,  \Nzero
\end{flalign}
where $Y^{z}\, =\, 0,\, Y^{\zb}\, =\, \frac{1}{\zb - \overline{w}}$. 
Hence although $O_{ab}$ has a trivial OPE with hard graviton modes, it must have non-trivial OPE with respect to $P(z)/P(\zb)\, =\, O_{\triangle=1, \pm 2}$.\\
That $P(z)$ is not a descendent primary with respect to ${\cal T}_{ab}$  was observed in \cite{dfh}. We refer the reader to section five in \cite{dfh} for details.\footnote{In \cite{barnichalgroid}  Barnich has put forward a different perspective on the non-centrally extended eBMS algebra. Namely to view it as a ``centrally extended algebroid" instead of an algebra with a 2-cocycle anomaly. However the implications of the Lie-algebroid ideas in quantum theory are not clear to the authors. More in detail, the presence of the 2-cocycle appears to be in tension with the fact that quantized supertranslation and superrotation charges are symmetries of the S-matrix. It is not clear to us, how reinterpreting eBMS algebra as a Lie algebroid alleviates this tension.} 

Thus we expect that in the celestial CFT, our results may be equivalent to choosing $t_{ab}$ as the stress tensor which is obtained by deforming ${\cal T}_{ab}$ by ${ \cal O}_{ab}$. 
\subsection{Towards gBMS algebra}\label{tgbmsa}
Our analysis so far has been restricted to the extended BMS algebra in which superrotations are generated by (local) conformal Killing vector fields on the celestial sphere. In this section, we offer a few remarks about the possibility of generalising the analysis to quantize the gBMS algebra.
\begin{itemize}
\item In a seminal paper \cite{pdp}, the soft charge associated to smooth superrotations $\textrm{Diff}(S^{2})$ was shown to be a shadow transform of the celestial stress tensor (this stress tensor is the soft charge for eBMS superrotations in the conformal primary basis). However a crucial ingredient in this analysis was to consider generic configurations of $T_{ab}$ which are meromorphic on the celestial sphere. Our constraint analysis assumed that $T_{ab}$ has poles only at infinity and hence to use the ideas of \cite{pdp}, we need to generalise our analysis to ensure that all meromorphic Geroch tensors are included in the radiative phase space.
\item To allow for smooth superrotations, we  need to reincorporate the celestial metric $q_{ab}$ as part of the kinematical phase space, as originally presented in \cite{cp}. In such case, the analogue of the constraints (\ref{june201}) are more involved. In particular, the constraint $F^2_a$ becomes
\begin{flalign} \label{divT}
\textrm{D}^{b}\, T_{ab}\, =\, 0
\end{flalign}
where $\textrm{D}_{a}$  is the covariant derivative of the celestial metric,  $\textrm{D}_a q_{bc}=0$. Thus, the 2d metric and Geroch tensor are no longer decoupled but are related through the non-linear constraint \eqref{divT}. 
\item  Once the sphere metric is included in the radiative phase space, it also contains a mode $p^{ab}$ conjugate to $q_{ab}$ that is constrained to be certain combination of soft news modes (we refer the reader to \cite{cp} for details). This is likely to make the constraint analysis far more intricate. 
\end{itemize}
An investigation of these constraints and the resulting Dirac brackets will be pursued elsewhere. 

\section*{Acknowledgements}
We are grateful to Rafael Flauger for drawing our attention to the issue of cocycle extension in eBMS algebra and for crucial discussions which led us to this project. We would also like to thank Anupam A H, Sujay Ashok, Sayali Bhatkar, Chandramouli Chowdhury, Laurent Freidel, Guzman Hernandez-Chifflet, Arpan Kundu, Ruchira Mishra, Prahar Mitra, Silvia Nagy, Olga Papadoulaki, Aneesh P.B., Javier Peraza, Siddharth Prabhu, Suvrat Raju, Biswajit Sahoo, Ashoke Sen and  Pushkal Srivastava  for valuable discussions on issues related to asymptotic symmetries in Quantum Gravity over the years. We are especially indebted to Dileep Jatkar for patiently explaining to us many facets of the BMS algebra in celestial CFT.\\
AL would like to thank the organizers of ``Celestial Holography" at PCTS (Princeton Center for Theoretical Science) where part of this work was presented. He would also like to thank Laura Donnay, Sabrina Pasterski and Andrea Puhm for discussions. MC acknowledges support from PEDECIBA and ANII grant FCE-1-2019-1-155865.

\appendix
\def\Ib{\overline{I}}
\def\Jb{\overline{J}}

\section{Derivation of Dirac brackets}\label{appA}
In this section, we present a derivation of the ``physical" radiative phase space Poisson brackets presented in section \ref{secfour}. The first step (see e.g. \cite{HT}) is to study the matrix with entries given by the Poisson brackets between constraints. In our case we have $4 \times 2 \, \times \infty$  constraints given in Eq. \eqref{june201}. We write the corresponding matrix as
\be \label{diracmatrix}
\begin{pmatrix}
\{F^I(z), F^J(w)\} & \{F^I(z), F^{\bar{J}}(w) \}  \\
\{F^{\bar{I}}(z), F^J(w)\} & \{F^{\bar{I}}(z), F^{\bar{J}}(w) \}
\end{pmatrix}
\ee
where $I,J,\Ib,\Jb=0,1,2,3$ are constraint indices that distinguish   holomorphic and anti-holomorphic components of the constraint functions. For instance, $F^0(z) \equiv F^0_{zz}$ and $F^{\overline{0}}(z) \equiv F^0_{\zb\zb}$. We can evaluate the various entries of this matrix by using the kinematical PBs presented in section \ref{sec4}. It is immediate to see that brackets of the same helicity type vanish,
\be
\{F^I(z), F^J(w)\} = \{F^{\bar{I}}(z), F^{\bar{J}}(w) \} = 0.
\ee
Evaluating the PBs between constraints with opposite helicity, we find the only non-vanishing ones are\footnote{The function $F^3_{z z}$ suffers from the type of ambiguities discussed in section \ref{hardPBsec} when evaluating PBs. We take the prescription $\{F^0_{zz} , F^3_{\wb \wb} \}= - X_{F^0_{zz}}(F^3_{\wb \wb})=- 2\delta^{(2)}(z,w) $ and  $\{F^1_{zz} , F^3_{\wb \wb} \}= \{F^3_{zz} , F^3_{\wb \wb} \}=0$.}
\be
\{F^0_{zz} , F^1_{\wb \wb} \} = \frac{1}{2} \Nzero(z) \delta^{(2)}(z,w), \quad \{F^0_{zz} , F^3_{\wb \wb} \}  = - 2\delta^{(2)}(z,w) , \quad 
\{F^1_{zz} , F^2_{\wb} \}  = -\frac{1}{2}\partial_z \delta^{(2)}(z,w),
\ee
together with their complex conjugates. In the notation of \eqref{diracmatrix}, we write them as the 4 by 4 matrix
\be \label{subD-matrix}
\{F^I(z), F^{\bar{J}}(w) \}  =  \F^{I \bar{J}}(z) \delta^{(2)}(z,w)
\ee
where
\be  \label{subDmatrix}
\F^{I \bar{J}}(z) = - \frac{1}{2}
\begin{pmatrix}
 0 & - \Nzero(z) & 0 & 4 \\
\Nzero(z) & 0 & \partial_z & 0 \\
0 & \partial_{\zb} & 0 & 0 \\
-4 & 0 & 0 & 0 \\
\end{pmatrix}.
\ee
 Similarly $\{F^{\Ib}(z), F^J(w) \}  =  \F^{\Ib J} \delta^{(2)}(z,w)$ with $\F^{\Ib J}$ the complex conjugate of \eqref{subDmatrix}.
 
 Provided we can invert the differential operator $\partial_z$, the matrices \eqref{subDmatrix} and  \eqref{diracmatrix}, are invertible, implying the constraints are second-class. We shall now proceed under this assumption and discuss later the subtleties that arise in situations where  $\partial_z$ cannot be inverted. 

The inverse of \eqref{diracmatrix} can be written as
\be
\begin{pmatrix}
\{F^I(z), F^J(w)\} & \{F^I(z), F^{\bar{J}}(w) \}  \\
\{F^{\bar{I}}(z), F^J(w)\} & \{F^{\bar{I}}(z), F^{\bar{J}}(w) \}
\end{pmatrix}^{-1} = \begin{pmatrix}
0 & \K_{I \overline{J}}(z)  \\
\K_{\overline{I} J}(z) & 0
\end{pmatrix} \delta^{(2)}(z,w) ,
\ee
where $\K_{\overline{I} J}$ is the operator-inverse of \eqref{subDmatrix}
\be
\F^{I \overline{J}} \K_{\overline{J} K} = \delta^I_K,
\ee
given by
\be\label{dirmat}
\K_{\overline{I} J}(z) = 
 -\frac{1}{2}\begin{pmatrix}
 0 & 0 & 0 & -1 \\
0 & 0 & 4 \partial^{-1}_{\zb} & 0 \\
0 & 4\partial^{-1}_{z} & 0 & \partial^{-1}_{z} \Nzero(z)  \\
1 & 0 & \Nzero(z)  \partial^{-1}_{\zb}  & 0 \\
\end{pmatrix},
\ee
where $\partial^{-1}_{z} \Nzero(z)$ and $\Nzero(z)  \partial^{-1}_{\zb}$ should be understood as  compositions of integral and multiplicative operators. Similarly  $\K_{I \Jb}$ is given by the complex conjugate of \eqref{dirmat}.

The Dirac bracket between two functions $f$ and $g$ is then given by
\be
\{f,g\}_\star = \{f,g\}+ \{f,g\}_\extra \label{defDB}
\ee
where 
\be \label{defextra}
 \{f,g\}_\extra =  \int d^2 w \{f,F^{I}(w) \} \K_{I \Jb}(w) \{g,F^{\Jb}(w) \} + (I \leftrightarrow \Ib, J \leftrightarrow \Jb)
\ee
We would like to compute the Dirac brackets between the elementary variables
\be \label{elementaryvar}
 N_{ab}(u, \hat{x}), \, \Nzero(\hat{x}), \, C(\hat{x}), \, \None_{ab}(\hat{x}),\, T_{ab}(\hat{x}) \, .
\ee
We  start by displaying the PBs of elementary  variables with the constraints. Given the kinematical PBs of section \ref{sec4}, one finds 
\be
\{C(z), F^0_{ww} \}  =  2 D^2_w G(z,w), \quad  \{C(z), F^1_{ww} \}  =  -2 D^2_w C(w) G(z,w),
\ee
\be
\{\None_{zz}, F^0_{\wb \wb} \}  =  \frac{1}{2}\Nzero(z) \delta^{(2)}(z,w), \quad   \{\None_{zz}, F^2_{\wb}\} = - \frac{1}{2}\partial_w \delta^{(2)}(z,w),
\ee
\be
\{T_{zz},  F^1_{\wb \wb} \} = - \frac{1}{2} \delta^{(2)}(z,w) , \quad  \{N_{zz}(u,z) , F^1_{\wb \wb} \}  = \frac{1}{2} \delta^{(2)}(z,w),
\ee
(together with their complex conjugates) with all remaining brackets being zero. We can now evaluate the extra contribution \eqref{defextra} and hence the Dirac bracket \eqref{defDB}  for all pairs of elementary variables \eqref{elementaryvar}.  We start by noticing that since  $\Nzero$ Poisson commutes with all constraints, its Dirac bracket with other functions will coincide with the kinematical one. In particular, the only non-zero Dirac bracket involving $\Nzero$ is 
\be
\{C(z) , \Nzero(w) \}_\star = G(z,w).
\ee
Another pair of brackets which is left unchanged is that of the news tensor with itself. The reason being that the news tensor has non-trivial brackets only with $F^{1}$, but the corresponding components of $\K_{I \Jb}$ vanish. Thus,
\be
\{N_{zz}(z,u), N_{\wb \wb}(u',w) \}_\star   = \frac{1}{2} \delta'(u-u') \delta^{(2)}(z,w).
\ee
For the remaining variables, one finds the non-trivial brackets involve the composition of derivative and  inverse-derivative operators. As we explain below, we will deal with such terms through the use of  projection operators defined on the space of functions,
\begin{flalign}
\begin{array}{lll}
O_{w}\, =\, \partial_{w}\, \partial_{w}^{-2}\, \partial_{w}\\
O_{\wb}\, =\, \partial_{\wb}\, \partial_{\wb}^{-2}\, \partial_{\wb}
\end{array}
\end{flalign}
The remaining non-trivial brackets are then written as 
\ba
 \{ C(z) , \None_{w w} \}_\star& =& 2 (1- O_{w} )   D^2_w C(w) G(z,w),  \\ 
   \{\None_{zz}, T_{\wb\wb} \}_\star &=& - \frac{1}{2}(1\, -\, O_{w}\, )\delta^{(2)}(z,w), \\
   \{ N_{zz}(z,u), \None_{\wb \wb } \}_\star & = & \frac{1}{2}\, O_{w}\, \delta^{(2)}(z,w) 
\ea
The reason we get $O_{w}$ or its orthogonal complement is the following.\\
Up to this stage, we had assumed $\partial_{w}$ was invertible and hence $O_{w}\, =\, \partial_{w}\partial^{-2}_{w}\partial_{w}=1$. We however note that if we  work with smeared variables, the nature of the smearing function would alter this conclusion. In particular, for an anti-holomorphic smearing, the $\partial_w$ derivatives would vanish. To allow for this possibility, we  work with the expressions above interpreting $\partial_{w}\partial^{-2}_{w}\partial_{w}$ as a symmetric operator that annihilates anti-holomorphic functions and such that is the identity on the space of smooth functions modulo anti-holomorphic kernel. We note that our choice of $O_{w}\, =\, \partial_{w}\partial_{w}^{-2} \partial_{w}$ 
instead of an operator such as $\partial_{w}^{-1}\, \partial_{w}$ ensures that it is symmetric with respect to its action on smearing of either functions. The latter however has  orthogonal kernel and co-kernel. Its kernel is the space of anti-holomorphic functions of ${\bf C}$ and it has a trivial cokernel.

Since the computation of Dirac brackets is subtle, We conclude the appendix by verifying the Dirac brackets satisfy the Jacobi identity. It is easy to see that the only non-trivial condition comes from  
\begin{flalign}
\{\, \{\, \None_{zz},\, \Nzero(w)\, \},\, C(x)\, \}_\star\, +\, \{\, \{\, C(x),\, \None_{zz}\, \},\, \Nzero(w)\, \}_\star\, +\,  \{\, \{\, \Nzero(w),\, C(x)\, \},\, \None_{zz}\, \}_\star\, \, =\, 0
\end{flalign}
Since $\{\, \None_{zz},\, \Nzero(w)\, \}_\star$ vanishes, the Jacobi identity will be satisfied if, 
\begin{flalign} \label{jacid}
\{\, \{\, C(x),\, \None_{zz}\, \}_\star,\, \Nzero(w)\, \}_\star\, =\, -\,  \{\, \{\, \Nzero(w),\, C(x)\, \}_\star,\, \None_{zz}\, \}_\star.
\end{flalign}
Computing each side of \eqref{jacid} we find
\ba
\{\, \{\, C(x),\, \None_{zz}\, \}_\star,\, \Nzero(w)\, \}_\star\,  &=& 2 (1- O_{z} ) \{    D^2_z C(z) G(x,z) , \Nzero(w)\, \}_\star\,  \\
 &=&  2 (1- O_{z} )   D^2_z G(z,w) G(x,z) 
\ea
\ba
-\,  \{\, \{\, \Nzero(w),\, C(x)\, \}_\star,\, \None_{zz}\, \}_\star & = & \{ G(x,w),\None_{zz}\, \}_\star \\
& = & \frac{1}{2}(1- O_{z} ) \frac{\delta G(x,w)}{\delta T_{\zb \zb}}.
\ea
The last derivative can be evaluated from the condition $\delta(\D G )=\delta\D G + \D \delta G=0$, together with the equation \eqref{deltaDCid} for $\delta \D$. One finds
\be
\frac{\delta G(x,w)}{\delta T_{\zb \zb}} = 4  D^2_z G(z,w) G(x,z) ,
\ee
from which \eqref{jacid} immediately follows.


\begin{thebibliography}{99}

\bibitem{bms1} 
  H.~Bondi, M.~G.~J.~van der Burg and A.~W.~K.~Metzner,
  ``Gravitational waves in general relativity. 7. Waves from axisymmetric isolated systems,''
  Proc.\ Roy.\ Soc.\ Lond.\ A {\bf 269}, 21 (1962).

\bibitem{bms2} 
  R.~K.~Sachs,
  ``Gravitational waves in general relativity. 8. Waves in asymptotically flat spacetimes,''
  Proc.\ Roy.\ Soc.\ Lond.\ A {\bf 270}, 103 (1962).

\bibitem{strom1}
A.~Strominger,
``On BMS Invariance of Gravitational Scattering,''
JHEP \textbf{07}, 152 (2014)

\bibitem{stromST}
T.~He, V.~Lysov, P.~Mitra and A.~Strominger,
``BMS supertranslations and Weinberg\textquoteright{}s soft graviton theorem,''
JHEP \textbf{05}, 151 (2015)

\bibitem{stromSR}
D.~Kapec, V.~Lysov, S.~Pasterski and A.~Strominger,
``Semiclassical Virasoro symmetry of the quantum gravity $ \mathcal{S}$-matrix,''
JHEP \textbf{08}, 058 (2014)

\bibitem{stromzhibo}
A.~Strominger and A.~Zhiboedov,
``Gravitational Memory, BMS Supertranslations and Soft Theorems,''
JHEP \textbf{01}, 086 (2016)



\bibitem{suvrat-flat}
A.~Laddha, S.~G.~Prabhu, S.~Raju and P.~Shrivastava,
``The Holographic Nature of Null Infinity,''
SciPost Phys. \textbf{10}, 041 (2021)

\bibitem{suvrat-ads}
S.~Raju,
``Is Holography Implicit in Canonical Gravity?,''
Int. J. Mod. Phys. D \textbf{28}, no.14, 1944011 (2019)
\bibitem{cordova}
C.~C\'ordova and S.~H.~Shao,
``Light-ray Operators and the BMS Algebra,''
Phys. Rev. D \textbf{98}, no.12, 125015 (2018)

\bibitem{freidelorbit}
W.~Donnelly, L.~Freidel, S.~F.~Moosavian and A.~J.~Speranza,
``Gravitational Edge Modes, Coadjoint Orbits, and Hydrodynamics,''
[arXiv:2012.10367 [hep-th]]

\bibitem{barnichruzz}
G.~barnich, R.~Ruzziconi,
``Co-adjoint representation of BMS group on Celestial Riemann Surfaces."
 JHEP, \textbf{79}, 2021 
 
\bibitem{kirilov}
 A.~A.~Kirilov,
 ``Lectures on the Orbit Method",
Graduate Studies in Mathematics, \textbf{64}; 2004; 408 pp;  

\bibitem{btebms}
G.~Barnich and C.~Troessaert,
``Symmetries of asymptotically flat 4 dimensional spacetimes at null infinity revisited,''
Phys. Rev. Lett. \textbf{105}, 111103 (2010)

\bibitem{clgbms}
M.~Campiglia and A.~Laddha,
``Asymptotic symmetries and subleading soft graviton theorem,''
Phys. Rev. D \textbf{90}, no.12, 124028 (2014)

\bibitem{btchargealgebra}
G.~Barnich and C.~Troessaert,
``BMS charge algebra,''
JHEP \textbf{12}, 105 (2011)


\bibitem{jackiw}
R.~Jackiw,
``Anomalies and Cocycles"
Comments, Nuclear Physics (1985) 

\bibitem{dfh}
J.~Distler, R.~Flauger and B.~Horn,
``Double-soft graviton amplitudes and the extended BMS charge algebra,''
JHEP \textbf{08}, 021 (2019)

\bibitem{aar}
Anupam.~A.~H, A.~Kundu, K.~Ray,
``Double Soft Graviton Theorems and BMS Symmetries",
Phys. Rev. D \textbf{97}, (2018)

\bibitem{ebmsope}
A.~Fotopoulos, S.~Stieberger, T.~R.~Taylor and B.~Zhu,
``Extended BMS Algebra of Celestial CFT'',
JHEP \textbf{03}, 130 (2020)


\bibitem{confsoft1}
L.~Donnay, A.~Puhm and A.~Strominger,
``Conformally Soft Photons and Gravitons'',
JHEP \textbf{01}, 184 (2019)

\bibitem{confsoft2}
T.~Adamo, L.~Mason and A.~Sharma,
``Celestial amplitudes and conformal soft theorems,''
Class. Quant. Grav. \textbf{36}, no.20, 205018 (2019)

\bibitem{confsoft3}
A.~Puhm,
``Conformally Soft Theorem in Gravity'',
JHEP \textbf{09}, 130 (2020)

\bibitem{shamik}
S.~Banerjee, S.~Ghosh and R.~Gonzo,
``BMS symmetry of celestial OPE,''
JHEP \textbf{04}, 130 (2020)

\bibitem{alfredo}
A.~Guevara, E.~Himwich, M.~Pate and A.~Strominger,
``Holographic Symmetry Algebras for Gauge Theory and Gravity,''
[arXiv:2103.03961 [hep-th]]


\bibitem{compere}
G.~Comp\`ere, A.~Fiorucci and R.~Ruzziconi,
``Superboost transitions, refraction memory and super-Lorentz charge algebra,''
JHEP \textbf{11}, 200 (2018)
[erratum: JHEP \textbf{04}, 172 (2020)]



\bibitem{cp}
M.~Campiglia and J.~Peraza,
``Generalized BMS charge algebra,''
Phys. Rev. D \textbf{101}, no.10, 104039 (2020)

\bibitem{wbms}
L.~Freidel, R.~Oliveri, D.~Pranzetti and S.~Speziale,
``The Weyl BMS group and Einstein's equations,''
[arXiv:2104.05793 [hep-th]].'

\bibitem{clpilot} 
  M.~Campiglia and A.~Laddha,
  ``New symmetries for the Gravitational S-matrix,''
  JHEP {\bf 1504}, 076 (2015)

\bibitem{geroch} 
  R.~Geroch,
  ``Asymptotic structure of spacetime,'' 
 in  \emph{Asymptotic structure of spacetime}, ed. L. Witten, Plenum,
New York (1976)

\bibitem{lambdabms}
G.~Comp\`ere, A.~Fiorucci and R.~Ruzziconi,
``The $\Lambda$-BMS$_4$ charge algebra,''
JHEP \textbf{10}, 205 (2020)

\bibitem{comperenichols}
G.~Comp\`ere and D.~A.~Nichols,
``Classical and Quantized General-Relativistic Angular Momentum,''
[arXiv:2103.17103 [gr-qc]]


\bibitem{as81} 
  A.~Ashtekar and M.~Streubel,
  ``Symplectic Geometry of Radiative Modes and Conserved Quantities at Null Infinity,''
  Proc.\ Roy.\ Soc.\ Lond.\ A {\bf 376}, 585 (1981)




\bibitem{ashoke} 
  A.~Laddha and A.~Sen,
  ``Logarithmic Terms in the Soft Expansion in Four Dimensions,''
  JHEP {\bf 1810}, 056 (2018)
  
  \bibitem{biswajit} 
  B.~Sahoo and A.~Sen,
  ``Classical and Quantum Results on Logarithmic Terms in the Soft Theorem in Four Dimensions,''
  JHEP {\bf 1902}, 086 (2019)


\bibitem{cl1903}
M.~Campiglia and A.~Laddha,
``Loop Corrected Soft Photon Theorem as a Ward Identity,''
JHEP \textbf{10}, 287 (2019)

\bibitem{sayali}
S.~Atul Bhatkar,
``Ward identity for loop level soft photon theorem for massless QED coupled to gravity,''
JHEP \textbf{10}, 110 (2020)



\bibitem{subsub2} 
  M.~Campiglia and A.~Laddha,
  ``Sub-subleading soft gravitons and large diffeomorphisms,''
  JHEP {\bf 1701}, 036 (2017)

\bibitem{pope}
H.~Godazgar, M.~Godazgar and C.~N.~Pope,
``New dual gravitational charges,''
Phys. Rev. D \textbf{99}, no.2, 024013 (2019)

\bibitem{celestialtorus}
A.~Atanasov, A.~Ball, W.~Melton, A.~M.~Raclariu and A.~Strominger,
``$(2,2)$ Scattering and the Celestial Torus,''

\bibitem{comperelong}
G.~Comp\`ere and J.~Long,
``Vacua of the gravitational field,''
JHEP \textbf{07}, 137 (2016)


\bibitem{sraction}
K.~Nguyen and J.~Salzer,
``The Effective Action of Superrotation Modes,''
[arXiv:2008.03321 [hep-th]].


\bibitem{btaspects} 
  G.~Barnich and C.~Troessaert,
  ``Aspects of the BMS/CFT correspondence,''
  JHEP {\bf 1005}, 062 (2010)

\bibitem{aalectures} 
  A.~Ashtekar,
  ``Asymptotic Quantization'',
  Naples, Italy: Bibliopolis (1987) 

\bibitem{stromIR}
D.~Kapec, M.~Perry, A.~M.~Raclariu and A.~Strominger,
``Infrared Divergences in QED, Revisited,''
Phys. Rev. D \textbf{96}, no.8, 085002 (2017)

\bibitem{akhoury1}
S.~Choi, U.~Kol and R.~Akhoury,
``Asymptotic Dynamics in Perturbative Quantum Gravity and BMS Supertranslations,''
JHEP \textbf{01}, 142 (2018)
\bibitem{akhoury2}
S.~Choi and R.~Akhoury,
``BMS Supertranslation Symmetry Implies Faddeev-Kulish Amplitudes,''
JHEP \textbf{02}, 171 (2018)

\bibitem{acl} 
  A.~Ashtekar, M.~Campiglia and A.~Laddha,
  ``Null infinity, the BMS group and infrared issues,''
  Gen.\ Rel.\ Grav.\  {\bf 50}, no. 11, 140 (2018)
  
\bibitem{pate}
E.~Himwich, S.~A.~Narayanan, M.~Pate, N.~Paul and A.~Strominger,
``The Soft $\mathcal{S}$-Matrix in Gravity,''
JHEP \textbf{09}, 129 (2020)

\bibitem{barnichalgroid}
G.~Barnich,
``Centrally Extended BMS4 Lie Algebroid", 
JHEP \textbf{06} 007 (2017)

\bibitem{pdp}
L.~Donnay, S.~Pasterski, A.~Puhm,
``Asymptotic Symmetries and Celestial CFT"
JHEP \textbf{09} 176 (2020)



\bibitem{HT}
M.~Henneaux and C.~Teitelboim,
``Quantization of gauge systems,''
Princeton, USA: Univ. Pr. (1992) 520 p
\end{thebibliography}
\end{document}